\documentclass[12pt]{article}

\usepackage{graphicx}
\usepackage{epstopdf}

\def\onehalf{{\textstyle \frac12}}
\def\ii{{\rm i}}
\def\dd{{\rm d}}
\def\rr{{\bf r}}

\def\ssr#1{{\scriptscriptstyle\rm #1}}
\def\of#1{{{\scriptstyle(}#1{\scriptstyle)}}}

\def\abs#1{{{\scriptstyle|}#1{\scriptstyle|}}}
\def\aabs#1{{{\scriptscriptstyle|}#1{\scriptscriptstyle|}}}

\def\tsty#1#2{{\textstyle\frac{#1}{#2}}}

\def\jour#1#2#3#4{{\it #1{}} {\bf #2}, #3 (#4)}
\def\lab#1{\label{eq:#1}}
\def\rf#1{(\ref{eq:#1})}
\def\Lie#1{\hbox{\sf #1}}

\newcommand{\be}{\begin{equation}}
\newcommand{\ee}{\end{equation}}
\newcommand{\bea}{\begin{eqnarray}}
\newcommand{\eea}{\end{eqnarray}}

\begin{document}

\begin{center}
{\Large\bf Interbasis expansions in the Zernike system}

\bigskip
	Natig M.\ Atakishiyev,\footnote{Instituto de Matem\'aticas,
	Universidad Nacional Aut\'onoma de M\'exico, Cuernavaca.}
     George S.\ Pogosyan,\footnote{Departamento de Matem\'aticas,
	Centro Universitario de Ciencias Exactas e Ingenier\'ias,
	Universidad de Guadalajara, M\'exico; Yerevan State University,
	Yerevan, Armenia; and Joint Institute for Nuclear Research,
	Dubna, Russian Federation.} \\
	 Kurt Bernardo Wolf,\footnote{Instituto de Ciencias F\'isicas,
	 Universidad Nacional Aut\'onoma de M\'exico, Cuernavaca.}
	 and Alexander Yakhno\footnote{Departamento de
	Matem\'aticas, Centro Universitario de Ciencias Exactas e
	Ingenier\'ias, Universidad de Guadalajara, M\'exico.}\\
	\bigskip\today

\end{center}


\begin{abstract}
  The differential equation with free boundary conditions on
  the unit disk that was proposed by Frits Zernike in 1934
  to find Jacobi polynomial solutions (indicated as I), serves to define a
  classical and a quantum system which have been found to be
  superintegrable. We have determined two new orthogonal polynomial
  solutions (indicated as II and III) that are separable, and which
  involve Legendre and Gegenbauer polynomials.
  Here we report on their three interbasis expansion coefficients:
  between the I--II and I--III bases they are given by $_3F_2(\cdots|1)$
  polynomials that are also special \Lie{su($2$)} Clebsch-Gordan 
  coefficients and Hahn polynomials. Between the II--III bases, we find
  an xpansion expressed by $_4F_3(\cdots|1)$'s and Racah polynomials  
  that are related to the Wigner $6j$ coefficients.

\end{abstract}

\section{Introduction}
				\lab{sec:one}

The two-dimensional differential equation proposed
by Frits Zernike in 1934 to find a basis of
orthogonal functions over the closed unit disk,
${\cal D}:=\{\rr=(x,y)\,|\,\abs\rr\le1\}$,
is \cite{Zernike34}
\be
      \widehat Z\,\psi\of\rr := \Big(\nabla^2
              -(\rr\cdot\nabla)^2
              -2\rr\cdot\nabla \Big)
                      \psi\of\rr = -E\, \psi\of\rr.
                      \lab{Zernikeq}
\ee
The operator $\widehat Z$ is hermitian under the
natural inner product of functions over this region,
\be
	(\psi_1,\psi_2)_{\cal D}:=\int_{\cal D} \dd^2\rr\,
		\psi_1\of\rr^*\,\psi_2\of\rr,    \lab{D-inn-prod}
\ee
that defines the space ${\cal L}^2({\cal D})$ of square-integrable
functions over $\cal D$. There,
$(\widehat Z \psi_1,\psi_2)_{\cal D}=(\psi_1,\widehat Z \psi_2)_{\cal D}$
and the eigenfunction solutions $\Psi_{n,m}\of\rr$ to \rf{Zernikeq}
will be orthogonal when they belong to different eigenvalues
$E_n=n(n+2)$, and/or different eigenvalues under the angular
momentum operator $\widehat M:=-\ii(x\partial_y-y\partial_x)$,
which corresponds to the evident rotational symmetry of
Zernike's equation and the ensuing separation of variables
in a polar coordinate system $\rr=(r,\phi)$.

Equation \rf{Zernikeq} can be seen as a classical 
Hamiltonian (with momenta ${\bf p}=-\ii\nabla$ 
as in Ref.\ \cite{PWY}), or a Schr\"odinger
equation with a non-standard quantum Hamiltonian
$\widehat H=-\onehalf\widehat Z$, as done in Ref.\ \cite{PSWY}.
In this paper we shall address the expansions between
the original Zernike eigenbasis (labelled I) and two of
the new separated eigenbases reported in \cite{PSWY}
(labelled II and III). All three eigenbases
are solutions of \rf{Zernikeq} that {\it separate\/}
into two polynomial factors:
for I, in Jacobi polynomials of the radius $r$ and
trigonometic functions of the angle $\phi$; for
II and III, the factors are Legendre and Gegenbauer
polynomials of coordinates that are not orthogonal
over the disk.

We consider to be relevant that the Zernike system is
one of the very few superintegrable systems that
have been actually used in optics, concretely
for phase-contrast microscopy \cite{Zernike34}. 
Superintegrability means that the system has more
constants of motion than degrees of freedom, 
that classically the system has closed orbits, and
Poisson brackets or commutators of the conserved
quantities or operators will generally
yield quadratic or higher elements of known
Lie algebras; see Refs.\ \cite{GPS-I}-\cite{MPW}.
As a quantum system, it is novel in the sense of
not being restricted in space by potential barriers,
but by boundaries where the wavefuctions can
have finite values and normal derivatives. Also,
as a Hamiltonian operator, we note that $\widehat Z$ is built
as linear combination of generators of an
\Lie{sp($2$,R)} algebra ($\sim\nabla^2$, $\rr\cdot\nabla$,
and $\abs\rr^2$) plus one quadratic term,
$(\rr\cdot\nabla)^2$.

\begin{figure}[t]
\centering
\includegraphics[scale=0.08]{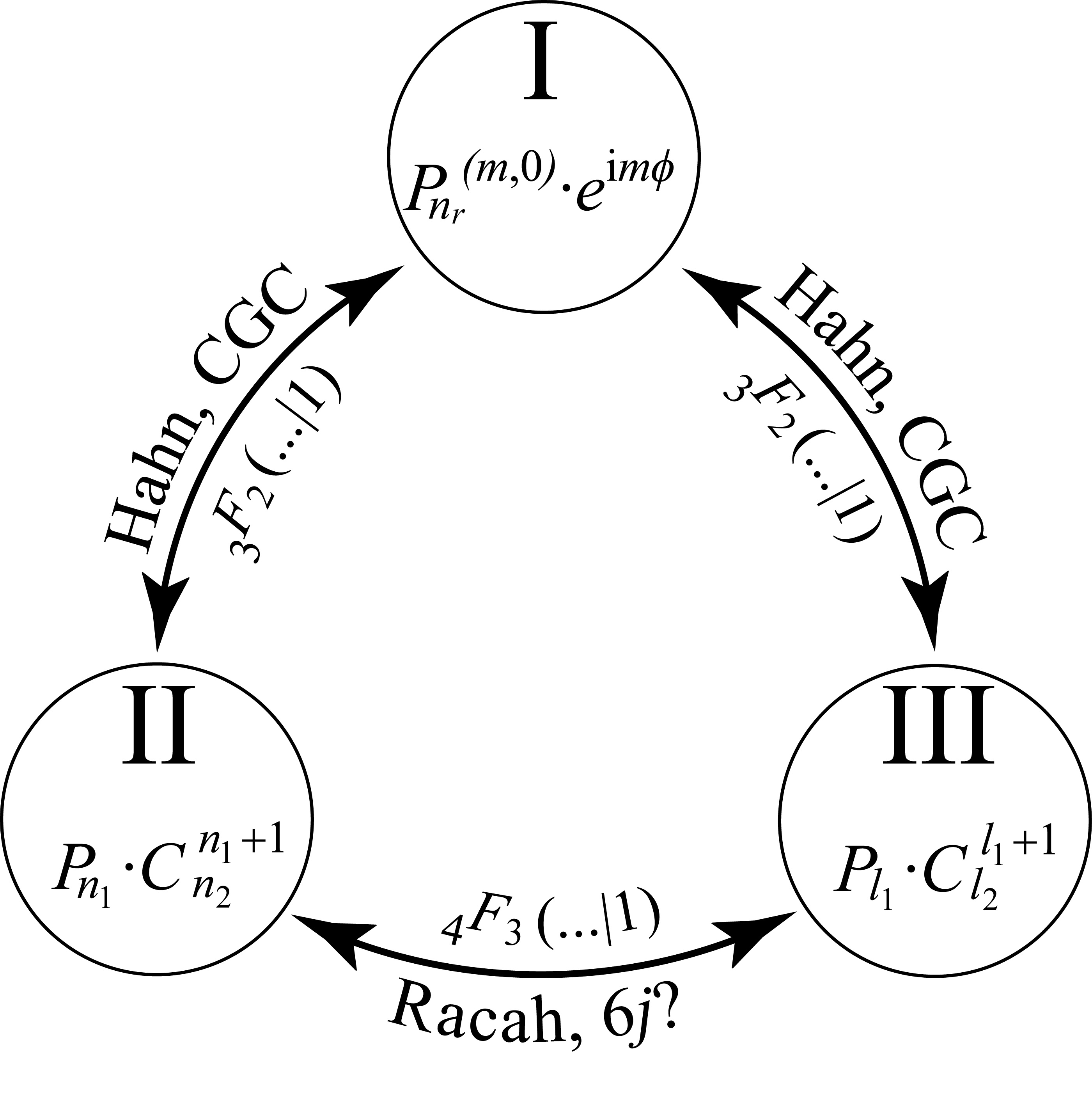}
\caption[]{Three bases and three interbasis expansions.
The original Zernike disk solutions \cite{Zernike34}, indicated I,
involve Jacobi polynomials and phases. These are related through 
interbasis expansions given by $_3F_2(\cdots|\,1)$'s, which are
Hahn polynomials and special Clebsch-Gordan coefficients,
to the new solutions, II and III \cite{PSWY}, 
that involve Legendre and Gegenbauer polynomials. 
The relation between the II and III bases are given by
$_4F_3(\cdots|\,1)$'s that are Racah polynomials, which
are suggested to be special $6j$ coupling coefficients.}
\label{fig:tres-relaciones}
\end{figure}

In Sect.\ \ref{sec:two} we succinctly recall the 
construction and expressions of the three said bases. Then
we proceed to find the interbasis expansions
I--II and I-III in Sect.\ \ref{sec:three}, which
yield special Clebsch-Gordan coefficients that
are Hahn polynomials, and  II-III in Sect.\ \ref{sec:four},
which are special $6j$ coefficients given by Racah polynomials,
as illustrated in Fig.\ \ref{fig:tres-relaciones}. 
We add  conclusions in Sect.\ \ref{sec:five}, while 
necessary but extensive derivations are collected in 
the Appendices.

\section{Three orthonormal eigenbases}
				\label{sec:two}
				
\begin{figure}[t]
\centering
\includegraphics[scale=0.60]{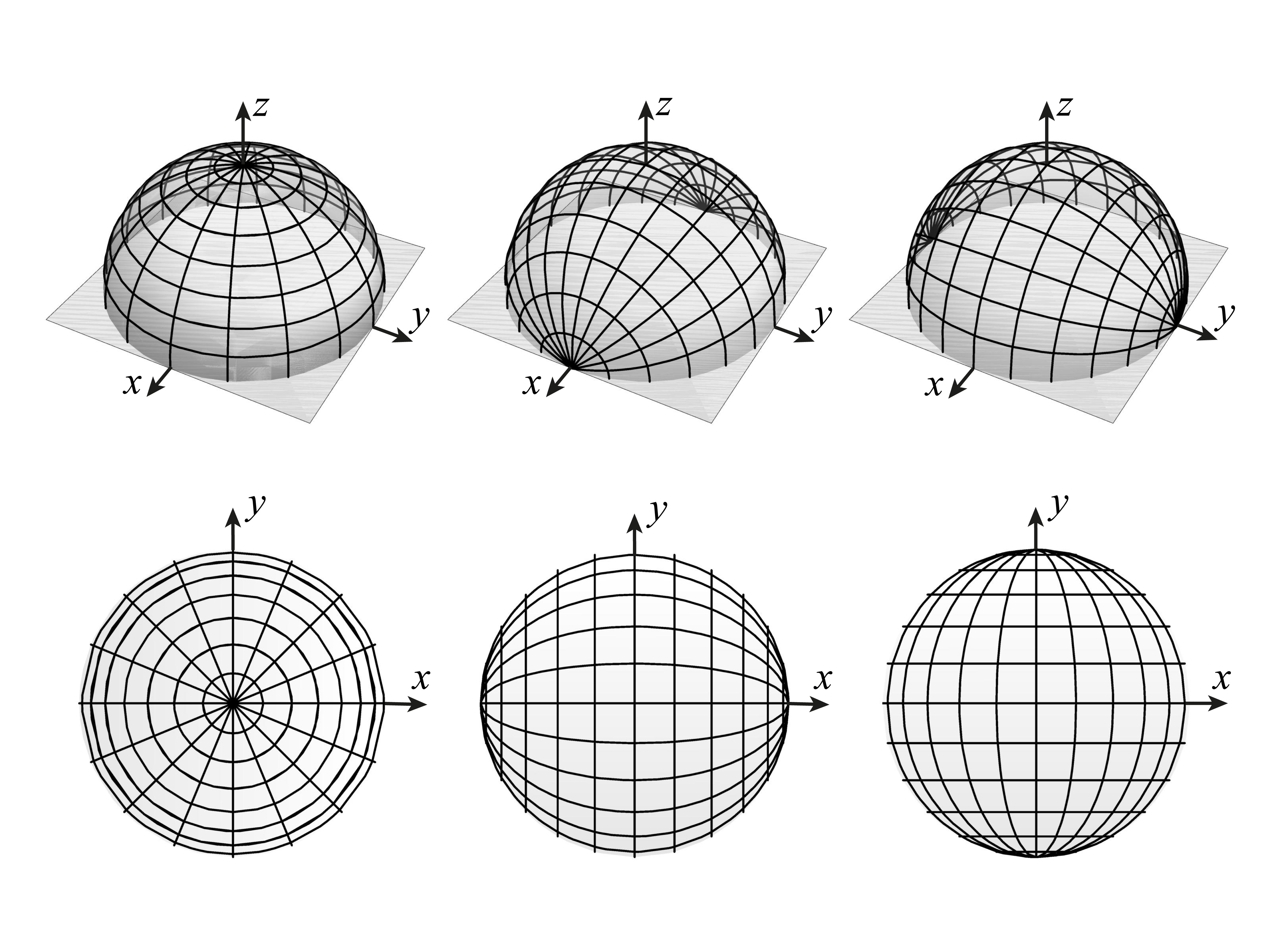}
\caption[]{{\it Top row\/}: The three coordinate systems
\rf{Syst-I}--\rf{Syst-III} on the half-sphere ${\cal H}_+$.
{\it Bottom row\/}: The same three coordinate systems
under projection to the disk $\cal D$.}
\label{fig:tresDH}
\end{figure}

The key to find new coordinate systems where the
solutions of Zernike's equation separate, is to
perform a vertical map from the disk $\cal D$ to
a half-sphere
${\cal H}_+:=\{\vec{r}=(x,y,z)\,|\,\abs{\vec{r}}=1,\,z\ge0\}$.
Separation of the solutions occurs when the coordinates
are such that one of them is constant on the boundary
of the region, i.e., on the circle $\abs\rr=1$ common
to both $\cal D$ and ${\cal H}_+$. As shown in Fig.\
\ref{fig:tresDH}, we can use the spherical coordinate
system $(\vartheta,\varphi)$ on the half-sphere, oriented
in three distinct directions. When the line of poles coincides
with the $z$-axis one obtains the polar separation of
coordinates of the Zernike basis I; when this line
coincides with the $x$- or $y$-axes, we obtain the
II or III bases respectively.

To find the coordinate ranges for the three spherical
coordinate systems, we introduce the unit 3-vector
$\vec{\bf\xi}$ of components
\be
	\xi_1:=x,\quad \xi_2:=y,\quad \xi_3:=\sqrt{1-x^2-y^2}\ge0.
		\lab{tres-xi}
\ee
The three coordinate systems and their ranges on ${\cal H}_+$ are 
defined as \cite{PWY,PSWY},
\bea
      &&\hskip-30pt\hbox{System I:} \lab{Syst-I} \\
              &&\hskip-30pt\xi_1 = \sin\vartheta\cos\varphi,\quad
              \xi_2=\sin\vartheta\sin\varphi,\quad
              \xi_3=\cos\vartheta,\qquad
              \vartheta|_0^{\pi/2},\ \varphi|_{-\pi}^{\pi}\,,
              \nonumber\\[5pt]
      &&\hskip-30pt\hbox{System II:} \lab{Syst-II} \\
              &&\hskip-30pt\xi_1 = \cos\vartheta',\quad
              \xi_2=\sin\vartheta'\cos\varphi',\quad
              \xi_3=\sin\vartheta'\sin\varphi',\qquad
              \vartheta'|_0^\pi,\ \varphi'|_0^\pi\,,
              \nonumber\\[5pt]
      &&\hskip-30pt\hbox{System III:} \lab{Syst-III} \\
              &&\hskip-30pt\xi_1 = \sin\vartheta''\sin\varphi'',\quad
              \xi_2=\cos\vartheta'',\quad
              \xi_3=\sin\vartheta''\cos\varphi'',\qquad
              \vartheta''|_0^\pi,\ \varphi''|_{-\pi/2}^{\pi/2}\,.
              \nonumber
\eea
The measure on ${\cal H}_+$ is related to that on $\cal D$ through
\be
	\dd^2S(\vec{\xi})=\frac{\dd\xi_1\,\dd\xi_2}{\xi_3}
		= \frac{\dd^2\rr}{\sqrt{1-\abs\rr^2}}
		= \sin\vartheta^\circ\,\dd\vartheta^\circ\,\dd\varphi^\circ,
			\lab{measures}
\ee
where $\vartheta^\circ\!\!,\varphi^\circ$ stands for any of the three
spherical coordinates in \rf{Syst-I}--\rf{Syst-III}, within their
corresponding ranges for the inner product on ${\cal H}_+$, 
the upper half-sphere, 
\be
	(\upsilon_1,\upsilon_2)_{{\cal H}_+}
		= \int_{{\cal H}_+}\dd^2S(\vec\xi)\,
		\upsilon_1(\vec\xi)^*\,\upsilon_2(\vec\xi)
		= (\psi_1,\psi_2)_{\cal D}.
			\lab{inn-prod-H}			
\ee
In accordance with the change of measures, the functions and the
Zernike operator on ${\cal H}_+$ and on $\cal D$ will then 
relate through
\be
	\upsilon(\vartheta^\circ\!,\varphi^\circ)\equiv	
	\upsilon(\vec\xi) := (1-\abs\rr)^{1/4}\psi\of\rr,\quad
	\widehat W := (1-\abs\rr)^{1/4}\widehat Z (1-\abs\rr)^{-1/4}.
		\lab{widehatW}
\ee

On $\vec\xi\in{\cal H}_+$, the Zernike differential
equation ---now with $\widehat W$--- has the simpler
Schr\"odinger structure
\be
	\widehat W\Upsilon(\vec\xi)=
		\Big(\Delta_\ssr{LB}+\frac{\xi^2_1+\xi^2_2}{4\xi^2_3}
			+ 1\Big)\Upsilon(\vec\xi)=-E\Upsilon(\vec\xi),
			\lab{W-diffeq}
\ee
where $\Delta_\ssr{LB}=L^2_1+L^2_2+L^2_3$ is the Laplace-Beltrami
operator on the sphere, with the formal \Lie{so($3$)} generators
$L_i:=\xi_j\partial_{\xi_k}-\xi_k\partial_{\xi_j}$ ($i,j,k$ cyclic),
and a repulsive oscillator-type of potential $\sim -r^2/(1-r^2)$
over the disk.

In Ref.\ \cite{PSWY} we wrote \rf{W-diffeq} in each of
the three coordinate systems \rf{Syst-I}--\rf{Syst-III},
separating each in successive or simultaneous differential
equations in $\vartheta^\circ$ and $\varphi^\circ$,
taking care that the solutions be square-integrable
under the inner product \rf{inn-prod-H}, and allowing them
to have finite values on the boundary circle.

\begin{figure}[t]
\centering
\includegraphics[scale=0.090]{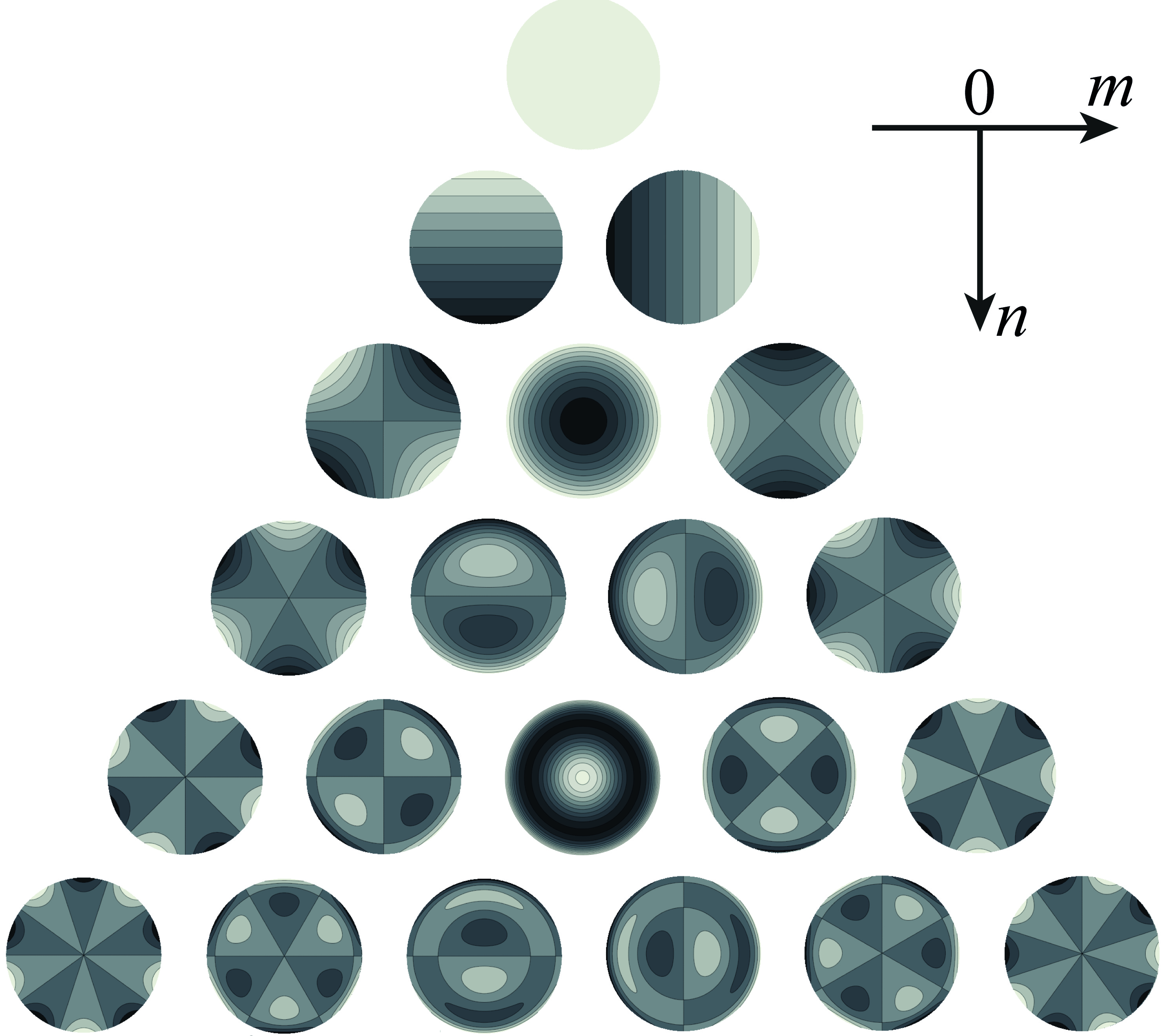}
\caption[]{The original Zernike solutions \cite{Zernike34} 
over the disk, $\Psi^\ssr{I}_{n,m}(r,\phi)$ in \rf{Psi-I}.
The rows are counted from $n=0$ down, and $m$ crosswise.
Since these functions are complex, 
$\Psi^\ssr{I}_{n,m}=\Psi^{\ssr{I}\,*}_{n,-m}$, 
we show their real part for $m\ge 0$ and their 
imaginary part for $m < 0$.}
\label{fig:Zern-I}
\end{figure}

\begin{figure}[t]
\centering
\includegraphics[scale=0.090]{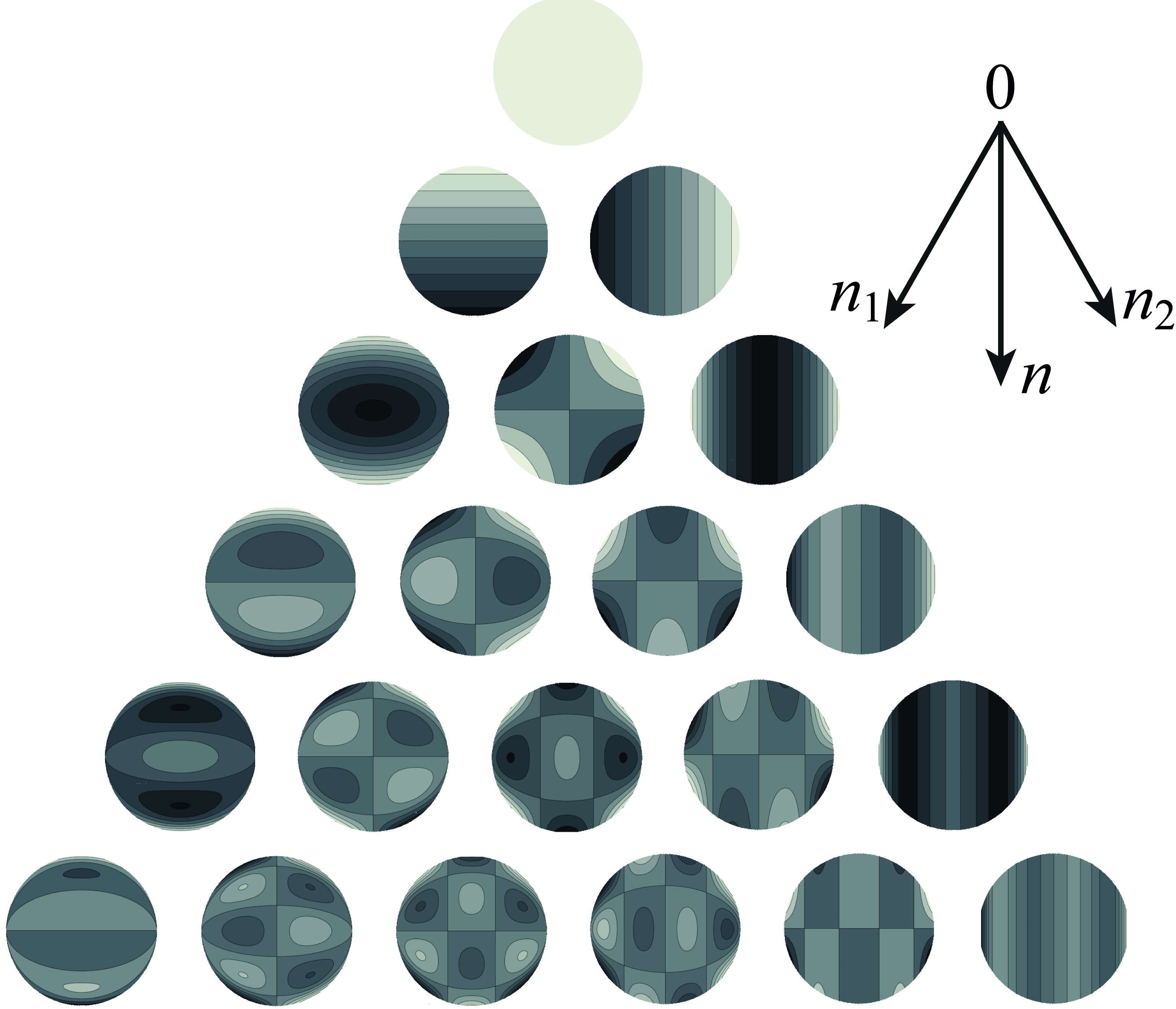}
\caption[]{The new solutions \cite{PSWY} of the Zernike equation
over the disk, $\Psi^\ssr{II}_{n_1,n_2}(x,y)$ in \rf{Psi-II}.
The rows are counted from $n=0$ down, $n_1$ counted down left
and $n_2$ down right.}
\label{fig:Zern-II}
\end{figure}

\begin{figure}[t]
\centering
\includegraphics[scale=0.78]{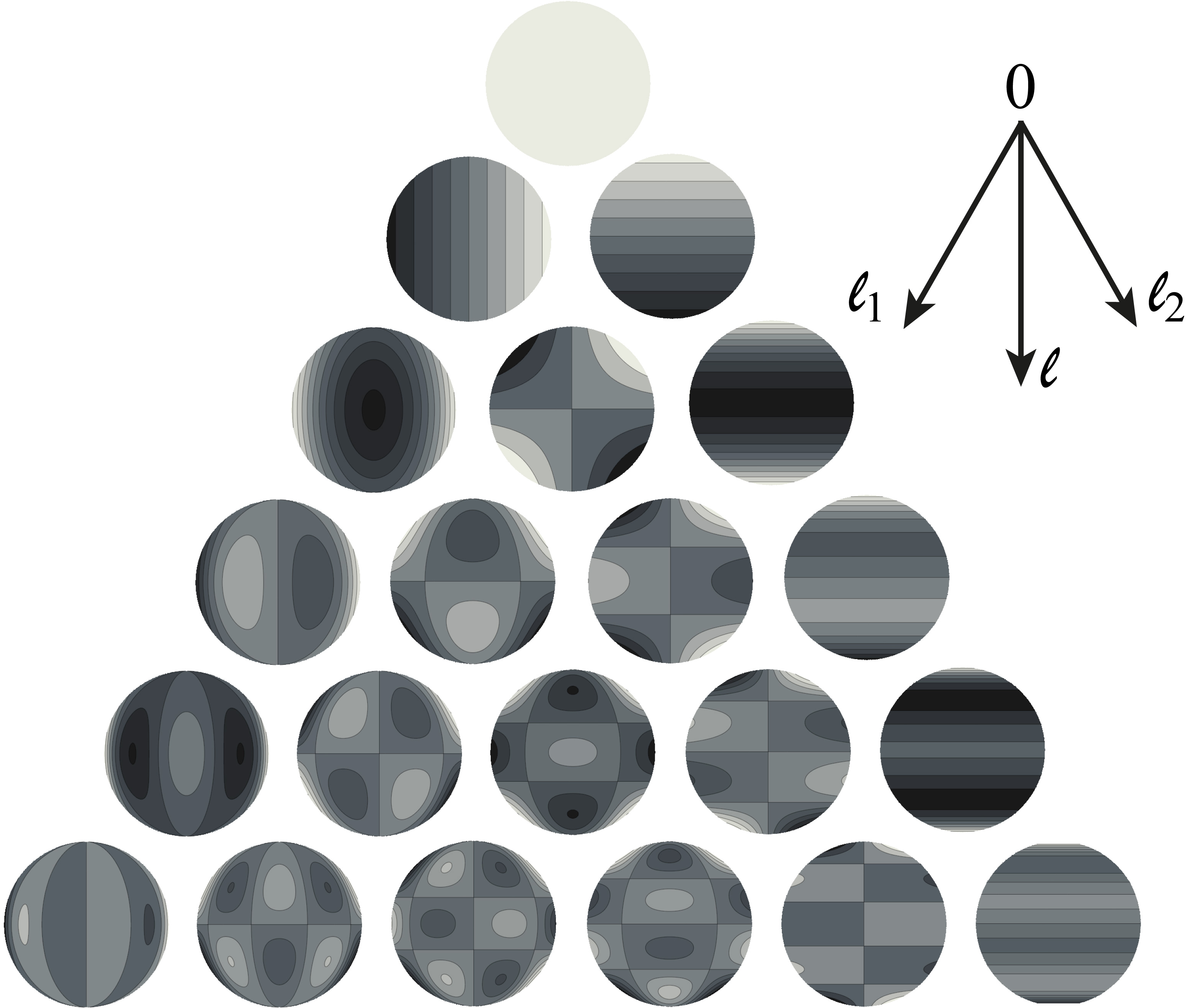}
\caption[]{The solutions $\Psi^\ssr{III}_{\ell_1,\ell_2}(x,y)$
of the Zernike equation defined in \rf{Psi-III} 
over the disk. The principal quantum number $\ell:=\ell_1+\ell_2$ 
here has the same interpretation as $n=n_1+n_2$ in the previous figure.}
\label{fig:Zern-III}
\end{figure}

The normalized eigen-solutions over ${\cal H}_+$
and $\cal D$, classified by `polar' $(n,m)$
and `Cartesian' $(n_1,n_2)$ eigenvalues are as follows:
\bea
\noalign{\noindent the original Zernike system I in
\rf{Syst-I}, shown in Fig.\ \ref{fig:Zern-I},}
	\Upsilon^\ssr{I}_{n,m}(\vartheta,\varphi)
	&=&\sqrt{\frac{n{+}1}{\pi}}\,
		(\sin\vartheta)^\aabs{m}(\cos\vartheta)^{1/2}
		P^{(\aabs{m},0)}_{\frac12(n-\aabs{m})}(\cos2\vartheta)\,
		e^{\ii m\varphi}, \lab{Ups-I}\\
	\Psi^\ssr{I}_{n,m}(r,\phi)
	&=&(-1)^{\frac12(n-\aabs{m})}\sqrt{\frac{n{+}1}{\pi}}\,
		r^\aabs{m}
		P^{(\aabs{m},0)}_{\frac12(n-\aabs{m})}(1{-}r^2)\,
		e^{\ii m\phi}; \lab{Psi-I}\\
\noalign{\noindent the $x$-oriented system II in
\rf{Syst-II}, shown in Fig.\ \ref{fig:Zern-II}, defined as}
	\Upsilon^\ssr{II}_{n_1,n_2}(\vartheta',\varphi')
	&=&C_{n_1,n_2}\,
		(\sin\vartheta')^{n_1+\frac12}C^{n_1+1}_{n_2}(\cos\vartheta')\,
		\sqrt{\sin\varphi'}P_{n_1}(\cos\varphi'),
		    \lab{Ups-II}\\
	\Psi^\ssr{II}_{n_1,n_2}(x,y)
	&=&C_{n_1,n_2}\,(1{-}x^2)^{\frac12 n_1}C^{n_1+1}_{n_2}(x)
		P_{n_1}\Big(\frac{y}{\sqrt{1{-}x^2}}\Big);
			\lab{Psi-II}\\
\noalign{\noindent and the $y$-oriented system III in \rf{Syst-III},
shown in Fig.\ \ref{fig:Zern-III}, defined as}
	\Upsilon^\ssr{III}_{\ell_1,\ell_2}(\vartheta'',\varphi'')
	&=&C_{\ell_1,\ell_2}\,
		(\sin\vartheta'')^{\ell_1+\frac12}C^{\ell_1+1}_{\ell_2}(\cos\vartheta'')\,
		\sqrt{\cos\varphi''}P_{\ell_1}(\sin\varphi''),	
	 		\lab{Ups-III}\\
	\Psi^\ssr{III}_{\ell_1,\ell_2}(x,y)
	&=&C_{\ell_1,\ell_2}\,P_{\ell_1}\Big(\frac{x}{\sqrt{1{-}y^2}}\Big)\,
	(1{-}y^2)^{\frac12 \ell_1}C^{\ell_1+1}_{\ell_2}(y),
		 \lab{Psi-III}\\
\noalign{\noindent where in \rf{Ups-II}--\rf{Psi-III} the multiplying
constant is}
	C_{k_1,k_2}&:=& 2^{k_1}\,k_1!\,\sqrt{\frac{(2k_1+1)(k_1+k_2+1)\,k_2!}{
					\pi(2k_1+k_2+1)!}}, \lab{constC}
\eea
where $J^\of{\alpha,\beta}_\nu$, $C^\alpha_\nu$ and $P_\nu$
are the Jacobi, Gegenbauer and Legendre polynomials of degree 
$\nu$ respectively,
and where $k_i\in \{0,1,2,\ldots\}=:{\cal Z}_0^+$ stands for
$n_i$ or $\ell_i$. The range of these quantum numbers is
\be
	\begin{array}{c}
	k_1+k_2= n = 2n_r+\abs{m} \in {\cal Z}_0^+,\\
	n_r\in{\cal Z}_0^+, \quad m\in\{-n,-n{+}2,\ldots n\}.
		\end{array}\lab{qmlabels}
\ee
Here $n=\ell$ is the {\it principal\/} quantum number
that determines the `energy' eigenvalues $E_n=n(n{+}2)$
in \rf{Zernikeq}; in system I, $n_r$ is the {\it radial\/} quantum number
(that counts the radial nodes), $m$ is the {\it angular\/}
quantum number, while $(k_1,k_2)$ in systems II and III
qualify to be called the {\it Cartesian\/} numbers.

It is important to remark that although the
multiplets formed with the quantum numbers
\rf{qmlabels} in Figs.\ \ref{fig:Zern-I}--\ref{fig:Zern-III}
exhibit the same pattern as
that of the two-dimensional quantum harmonic
oscillator in polar and Cartesian coordinates respectively,
this analogy is misleading. There is no Lie
algebra of raising and lowering operators
to provide two-term transitions within the
multiplet, only {\it three\/}-term differential
and recursion relations have been found and are
analyzed in several mathematical papers
\cite{Bhatia-Wolf,Born,Ismail,Kintner,Myrick,Shakibaei,Wunsche}. 
The constant of motion operators that one collects
when separating the solutions, form instead a
superintegrable cubic Higgs algebra \cite{Higgs} 
as shown in Ref.\ \cite{PSWY}.


\section{I-II and I-III Interbasis expansions}
				\label{sec:three}

We first consider the interbasis expansion between the I and II
orthonormal basis functions in \rf{Ups-I} and \rf{Ups-II}
for fixed values of the principal quantum number $n$,
\be
	\Upsilon^\ssr{II}_{n_1,n_2} (\vartheta', \varphi')
		= \sum_{m=-n\,(2)}^{n} W_{n_1, n_2}^{n,m}
			\Upsilon^\ssr{I}_{n,m}(\vartheta, \varphi), \lab{0-SOL1}
\ee
where $\sum_{m=-n\,(2)}^n$ indicates that the sum over $m$
is in the range \rf{qmlabels}, so the radial quantum number
$n_r=\onehalf(n-\abs{m})\in{\cal Z}_0^+$ is integer. The relation
between the primed and unprimed angles is found equating
\rf{Syst-I} and \rf{Syst-II}, as
\be
	\cos\vartheta' = \sin\vartheta \cos\varphi, \quad
	\cos\varphi' = \frac{\sin\vartheta \sin\varphi}{
			\sqrt{1-\sin^2\vartheta \cos^2\varphi}}.
				\lab{angles-II-I}
\ee

\subsection{I--II interbasis with $_3F_2$'s}

To calculate the explicit form of the expansion coefficients
$W_{n_1, n_2}^{n, m}$ it is sufficient to fix the value
of the coordinate that is constant over the boundary, and
then use the orthogonality of the wave functions for the
other coordinate in the expansion \rf{0-SOL1}.
Thus consider $\vartheta = \onehalf\pi-\varepsilon$ for
vanishing $\varepsilon$; then
$\cos\vartheta=-\sin\varepsilon\approx -\varepsilon$
and $\sin\vartheta=\cos\varepsilon\approx1$, and also
$P_{n_1}(1) = 1$ and $P_{n_r}^{(\aabs{m}, 0)}(-1) = (-1)^{n_r}$.
Dividing by a common vanishing factor, the expansion
\rf{0-SOL1} remains in terms only of functions of $\varphi$,
and reads
\be
	C_{n_1,n_2} \sqrt{\frac{\pi}{n{+}1}}
		(\sin\varphi)^{n_1}\, C^{n_1+1}_{n_2}(\cos\varphi)
		=\sum_{m=-n\,(2)}^{n} (-1)^{\frac12(n-\aabs{m})}
			W_{n_1, n_2}^{n,m}\, e^{\ii m \varphi}.
				\lab{expan2}
\ee
Thus, using the orthogonality of the functions $e^{\ii m \varphi}$
in $m$, we obtain the following integral representation
for the interbasis expansion coefficients,
\be
	W^{n, m }_{n_1, n_2} = \frac{(-1)^{\frac12(n-\aabs{m})}
		C_{n_1,n_2}}{2\sqrt{\pi(n+1)}}\,
		\int_{-\pi}^{\pi} \dd\varphi\,(\sin\varphi)^{n_1}\,
		C_{n_2}^{n_1+1}(\cos\varphi) \, e^{- \ii m \varphi}.
			\lab{INTER-01}
\ee
This integral does not appear in the standard tables, so
we perform its computation writing the functions in the
integrand as finite series in $e^{\ii\varphi}$,
\bea
	\sin^{k} \varphi
	&=& \frac{e^{\ii k \varphi}}{(2\ii)^{k}}\, (1-e^{-2\ii\varphi })^{k}=
	\frac1{(2\ii)^k}  \sum\limits_{l=0}^{k} \frac{(-1)^l\, k!}{l! \, (k-l)! }\,
		e^{\ii(k-2 l)\varphi}, \lab{APP-2-01}\\
	C^{\lambda}_{\mu}(\cos\varphi)
	&=& \sum_{j=0}^{\mu} \frac{\Gamma(\lambda +j)}{j! \, (\mu - j)!}\,
	\frac{\Gamma(\lambda + \mu - j)}{\Gamma^2(\lambda)}\, e^{-\ii(\mu-2j)\varphi}.\lab{APP-3-01}
\eea
Substituting these  expansions in \rf{INTER-01}, replacing
the constant $C_{n_1,n_2}$ from \rf{constC}, and using the orthogonality
of the $e^{\ii m\varphi}$ functions, we obtain a hypergeometric
$_3F_2$ terminating series of unit argument, 
\bea
	W^{n, m }_{n_1, n_2}
	&=& \frac{\ii^{n_1} (-1)^{\frac12(m+\aabs{m})} \, n!
		}{\Big(\frac12(n_1-n_2 - m)\Big)! \Big(\frac12(n + m)\Big)!}
		\, \sqrt{\frac{2n_1+1}{n_2!\,(n+ n_1+1)!}} \nonumber\\
	&&{\quad}\times {_3F_2}
		\bigg(\begin{array}{c} -n_2, \quad n_1 + 1,\quad  -\onehalf(n+m) \\
			-n,\quad  \onehalf(n_1-n_2 - m)+ 1 \end{array}\bigg|\,1\bigg),
		\lab{COEFFICIENT-01}
\eea
where we note that of the five parameters in $_3F_2(\cdots|1)$,
only three, e.g., $(n_1,n_2,m)$, are effectively present in the
interbasis coefficients.

\subsection{I--II interbasis with Clebsch-Gordan coefficients}  \label{sec:Clebsch}

	Perhaps surprinsingly, the interbasis coefficients 
$W^{n, m}_{n_1, n_2}$ in \rf{COEFFICIENT-01} can be 
compactly expressed in terms of \Lie{su($2$)} Clebsch-Gordan 
coefficients $C_{a, \alpha; b, \beta}^{c, \gamma}$ of a special type.
As given by Varshalovich {\it et al}.\ \cite{V}, the 
generic coefficients that couple angular 
momentum states $|a,\alpha\rangle$ and $|b,\beta\rangle$ to 
form $|c,\gamma\rangle$, after a transformation between 
two $_3F_2$ forms, are
\be
	\begin{array}{l}  \displaystyle
		C_{a,\alpha; b, \beta}^{c, \gamma}
			= \sqrt{\frac{(2c{+}1)(b{+}c{-}a)!(b{-}\beta)!(c{+}\gamma)!(c{-}\gamma)!
			  }{(a{+}b{-}c)!(a{-}b{+}c)!(a{+}b{+}c{+}1)!(a{+}\alpha)!(a{-}\alpha)!(b{+}\beta)!}}
			\\[5mm] \displaystyle
		{\ }\times \frac{\delta_{\gamma,\alpha{+}\beta}(2a)!\,(c{-}b{+}\alpha)!
			}{(c{-}b{+}\alpha)!\,(c{-}a{-}\beta)!}\,
		{_3F_2}\bigg( \begin{array}{c}
		{-}a{-}b{+}c,\ {-}a{+}\alpha,\  b{-}a{+}c{+}1\\
		{-}2a,\quad c{-}a{-}\beta{+}1\end{array}\bigg|\, 1\bigg).
		  \end{array}  \lab{CG2}
\ee
Now, comparing this with \rf{COEFFICIENT-01}, we
can write the $W^{n, m}_{n_1, n_2}$  coefficients, 
with  $a=b=\frac12 n$, $\alpha=-\beta=-\frac12 m$, and 
$\gamma=0$, in terms of a special 
type of Clebsch-Gordan coefficients and a phase, as
\be
	W^{n, m}_{n_1, n_2}= \ii^{n_1} (-1)^{\frac12(m+\aabs{m})}\,
		C_{\frac12{n}, \, -\frac12{m}; \, \frac12{n},\,\frac12{m}}^{n_1, \, 0}.
		 \lab{COEF2}
\ee

Let us note that values of 
$C_{\frac12{n}, \, -\frac12{m}; \, \frac12{n},\,\frac12{m}}^{n_1, \, 0}$ 
satisfy all necessary conditions \cite[\S\ 8.1.1]{V} for valid \Lie{su($2$)}
Clebsch-Gordan coefficients, namely:
the triangle condition because $0 \le n_1 \le n = n_1+n_2$,
with integer or half-integer non-negative numbers
$\abs{m} \le n$, $0\le n$.
Finally, we note that while the original Zernike solutions are complex,
$\Upsilon^\ssr{I}_{n,m}=\Upsilon^{\ssr{I}\,*}_{n,-m}$, the new ones,
$\Upsilon^\ssr{II}_{n_1,n_2}$ and $\Upsilon^\ssr{III}_{\ell_1,\ell_2}$ are real. 
This property is assured by relation between $\pm m$ coefficients,
\be
   C_{\frac12{n}, \, \frac12{m}; \, \frac12{n},\,-\frac12{m}}^{n_1, \, 0}
   =(-1)^{n_2}C_{\frac12{n}, \, -\frac12{m}; \, \frac12{n},\,\frac12{m}}^{n_1, \, 0}.
       \lab{CGC-real-imag}
\ee

The expansion inverse to \rf{0-SOL1}, namely
\be
	\Upsilon^\ssr{I}_{n,m}(\vartheta, \varphi)=
			\sum_{n_1 = 0}^{n} {\widetilde{W}}_{n, m}^{n_1, n_2}
		\Upsilon^\ssr{II}_{n_1,n_2} (\vartheta', \varphi'), \lab{EXP2}
\ee
with $n_1+n_2=n$, follows from the orthogonality property of 
the $\Lie{su($2$)}$ Clebsch-Gordan coefficients. These II-I 
interbasis coefficients are thus given by
\be
	\widetilde{W}_{n, m}^{n_1, n_2}= (-\ii)^{n_1} (-1)^{\frac12(m+\aabs{m})}\,
		C_{\frac12{n}, \, -\frac12{m}; \, \frac12{n}, \, \frac12{m}}^{n_1, \, 0},
			\lab{COEF3}
\ee
and may be written in terms of $_3F_2$ 
hypergeometric functions through \rf{CG2}.

\subsection{I--II interbasis with Hahn polynomials}

	The interbasis coefficients $W^{n, m}_{n_1, n_2}$ can be also
expressed in terms of the $N$ Hahn polynomials of degree $p$ of 
a discrete variable $x$ \cite{Koekoek:2010},
\be
	Q_{p}(x;\, \alpha, \beta, N)
	:= {_3F_2}	\bigg(\begin{array}{c} -p,\quad -x,\quad  p+\alpha+\beta+1 \\
			-N,\quad  \alpha+1 \end{array}\bigg|\,1\bigg), \lab{Hahn1}
\ee
for $x,p\in\{0,1,\ldots, N\}$. Applying the transformation
\be
	{_3F_2}\bigg( \begin{array}{c}
		a,\quad b,\quad c\\
		d,\quad e\end{array}\bigg|\, 1\bigg)
			=\frac{(d{+}1)!\, (d{-}a{-}b{+}1)!
				}{(d{-}a{+}1)!\,(d{-}b{+}1)!}\,
	{_3F_2}\bigg( \begin{array}{c}
		a,\quad b,\quad e{-}c\\
		a{+}b{-}d{+}1,\quad e\end{array}\bigg|\, 1\bigg), \lab{F32_1}
\ee
to the ${_3F_2}$ hypergeometric polynomial in \rf{COEFFICIENT-01},
we can write the interbasis coefficients $W^{n, m }_{n_1, n_2}$,
with its three effective parameters, as
\be
	\begin{array}{l}  \displaystyle
		W^{n, m }_{n_1, n_2}  = \frac{ \ii^{n_1}\,(-1)^{\frac12(m+\aabs{m})} (n!)^2
			}{ \Big(\frac12(n - m)\Big)!\, \Big(\frac12(n + m) \Big)! } \,
				\sqrt{\frac{2 n_1 +1}{n_2!\, (n+n_1+1)!}}\\[5pt] \displaystyle
		{\qquad\qquad}\times Q_{n_2}\Big(\onehalf(n+m);\, -n-1,\, -n-1,\, n\Big).
			\end{array}	\lab{COEF12}
\ee

The discrete orthogonality relation for the Hahn polynomials is
of the form \cite[Eq.\ (9.5.2)]{Koekoek:2010}
\be
   \sum_{j=0}^{N}\,\rho(j)\,Q_m(j;\alpha,\beta,N)\,Q_n(j;\alpha,\beta,N)
   = \delta_{m,n}\,d_n^2, \lab{orthoHahn}
\ee
with the weight function $\rho(j)$ and the norm $d_n$, 
\be
   \rho(j) = \bigg( \frac{N!}{j!\,(N-j)!}\bigg)^2, \qquad
   d_n = \frac{1}{N!}\,\sqrt{\frac{n!\,(2N + 1-n)!}{2N-2n + 1}}.
   \lab{rho-d}
\ee
Thus \rf{COEF12} can be written in a more compact form as
\be
   W^{n, m }_{n_1, n_2}  = \ii^{n_1}\,(-1)^{\frac12(m+\aabs{m})}
   \frac{\sqrt {\rho (\frac12\of{n{+}m})}}{d_{n_2}}\,
   Q_{n_2}\left(\onehalf(n{+}m);\, -(n{+}1), -(n{+}1), n\right). \lab{COEF12'}
\ee

In fact, the Hahn polynomials present here are particular cases of
\rf{Hahn1}, with $\alpha = \beta = -(n{+}1) < -N$ and $N = n$, and
symmetric under $m\leftrightarrow {-m}$, which coincide with
the {\it dual\/} Hahn polynomials
\cite[Eq.\ (9.5.2)]{Koekoek:2010},
\be
	R_{\frac12(n+m)} \Big(\lambda(n_2);\, -(n{+}1),-(n{+}1), n\Big)
		=Q_{n_2}\Big(\onehalf(n+m);\, -(n{+}1),-(n{+}1), n\Big),
		\lab{dual-H}
\ee
on the quadratic lattice $\lambda(n_2):=n_2(n_2 - 2n -1)$
(see remark in \cite[p.\ 208]{Koekoek:2010}).
The expansion inverse to \rf{0-SOL1}, namely \rf{EXP2},
follows from the orthogonality of Hahn and dual Hahn 
polnomials given in \cite[Eq.\ (9.6.2)]{Koekoek:2010}.

\subsection{The I--III interbasis expansion}

The coefficients of the interbasis expansion between the I and 
III bases can be found with the same method as for the I-II 
interbasis coefficients \rf{COEFFICIENT-01},
\rf{COEF12'}, or \rf{COEF2} given above, by realizing that the
spherical coordinates $(\vartheta',\varphi')$ in \rf{Syst-II} and
\rf{Syst-III} are related through $\vartheta'\mapsto\vartheta''$
and $\varphi'\mapsto\varphi''+\frac12\pi$, and up to a phase
$(-1)^{\ell_1}$. The expansion between
the solutions defined in \rf{Ups-I} and \rf{Ups-III} is 
\be
	\Upsilon^\ssr{III}_{\ell_1,\ell_2} (\vartheta'', \varphi'')
		= \sum_{m=-n\,(2)}^{n} \widehat W_{\ell_1, \ell_2}^{n,m}\,
       \Upsilon^\ssr{I}_{n,m} (\vartheta, \varphi),
           \lab{inverse-exp}
\ee
with the relation between the angles being now
\be
   \cos\vartheta''=\sin\vartheta\sin\varphi, \quad
   \cos\varphi''=\frac{\cos\vartheta}{\sqrt{1-\sin^2\vartheta\sin^2\varphi}}.
       \lab{inv-angles-I-III}
\ee

To find the coefficients $\widehat W_{\ell_1, \ell_2}^{n,m}$,
one comes to an integral similar to \rf{INTER-01} except
for the trigonometric functions of $\varphi$, namely
\be
	\widehat{W}_{\ell_1, \ell_2}^{n,m} = \frac{(-1)^{\frac12(n-\aabs{m})}
		C_{\ell_1,\ell_2}}{2\sqrt{\pi(n+1)}}\,
		\int_{-\pi}^{\pi} \dd\varphi\,(\cos\varphi)^{\ell_1}\,
		C_{\ell_2}^{\ell_1+1}(\sin\varphi) \, e^{- \ii m \varphi},
			\lab{INTER-III-I_1}
\ee
so with the change of variables $\varphi \to \varphi + \onehalf\pi$ 
and the same procedure used in \rf{INTER-01}, we obtain
\be
	\widehat{W}_{\ell_1, \ell_2}^{n,m} 
	= (-1)^{\ell_1} \exp(-\ii \onehalf\pi m)\, 
	W_{\ell_1, \ell_2}^{n,m},
\ee
where the coefficients $W_{\ell_1, \ell_2}^{n,m}$ are those in 
\rf{0-SOL1} and \rf{COEFFICIENT-01}, with $(\ell_1,\ell_2)$ 
replacing $(n_1,n_2)$.


\section{II-III interbasis expansions}   \label{sec:four}

	We consider now the interbasis expansion between the
two new spherical wave functions, 
$\Upsilon^\ssr{II}_{n_1,n_2}(\vartheta',\varphi')$ defined in \rf{Ups-II} and
$\Upsilon^\ssr{III}_{\ell_1,\ell_2}(\vartheta'',\varphi'')$ in \rf{Ups-III},
within the same multiplet characterized by the principal 
quantum number $n$ that contains $n+1$ functions,
\be
	\Upsilon^\ssr{III}_{\ell_1,\ell_2}(\vartheta'', \varphi'')
		= \sum^{n}_{n_2=0} U_{\ell_1, \ell_2}^{n_1, n_2}\,
		\Upsilon^\ssr{II}_{n_1,n_2}(\vartheta', \varphi'),
		\lab{0-SOL1_II}
\ee
where $\ell_1+\ell_2=n=n_1+n_2$, $\ell_i,n_i\in{\cal Z}_0^+$,
and with the relation between the two spherical coordinate systems
$(\vartheta'', \varphi'')$ and $(\vartheta', \varphi')$ being now
\be
	\begin{array}{rl}
	\cos\vartheta'' = \sin\vartheta' \cos\varphi', &
	\sin\vartheta'' = {\sqrt{1-\sin^2\vartheta' \cos^2\varphi'}},\\[5pt]
	\displaystyle \cos\varphi'' =
		\frac{\sin\vartheta' \sin\varphi'}{\sqrt{1-\sin^2\vartheta' \cos^2\varphi'}},&
	\displaystyle\sin\varphi'' =
		\frac{\cos\vartheta'}{\sqrt{1-\sin^2\vartheta' \cos^2\varphi'}}.
			\end{array} \lab{COOR-01}
\ee

The interbasis expansion coefficients in \rf{0-SOL1_II} 
can be succinctly expressed by passing through 
$\Upsilon^\ssr{I}_{n,m}(\vartheta, \varphi)$, 
using the coefficients for the inverse expansion 
in \rf{inverse-exp} and the direct one in \rf{EXP2}, as
\be
	\Upsilon^\ssr{III}_{\ell_1,\ell_2}(\vartheta'', \varphi'')
		=\sum_{m=-n\,(2)}^{n} \widehat{W}_{\ell_1, \ell_2}^{n,m}\, 
   \sum_{n_1 = 0}^{n} {\widetilde{W}}_{n, m}^{n_1, n_2}\,
		\Upsilon^\ssr{II}_{n_1,n_2} (\vartheta', \varphi').
           \lab{Sumsum}
\ee
Replacing now the Clebsch-Gordan coefficients from 
\rf{COEF2} with care of the phases, we find
\be
   U_{\ell_1, \ell_2}^{n_1, n_2}=(-1)^{\ell_1}\sum_{m=-n\,(2)}^{n} \ii^{\ell_1-n_1-m} 
       C_{\frac12{n}, \, -\frac12{m}; \, \frac12{n}, \, \frac12{m}}^{\ell_1, \, 0}\,
       C_{\frac12{n}, \, -\frac12{m}; \, \frac12{n}, \, \frac12{m}}^{n_1, \, 0},
           \lab{U-CC}
\ee
and changing the summation index $m$ to $k=\onehalf(m+n)\in\{0,1,\ldots,n\}$
this expression is rewritten as 
\be
   U_{\ell_1, \ell_2}^{n_1, n_2}= \ii^{\ell_1+n_2} \sum_{k=0}^{n} (-1)^{\ell_1+k} 
       C_{\frac12{n}, \, \frac12{n}-k; \, \frac12{n}, \,  -\frac12{n} + k}^{\ell_1, \, 0}\,
       C_{\frac12{n}, \, \frac12{n}-k; \, \frac12{n}, \,  -\frac12{n} + k}^{n_1, \, 0}.
           \lab{U-CC2}
\ee
As will be seen in the next subsection, when $\ell_1+n_2$ or $\ell_2+n_1$  
are odd numbers, the coefficients $U_{\ell_1, \ell_2}^{n_1, n_2}$ 
are zero; the non-zero coefficients are real, 
as are the two sets of basis functions 
$\Upsilon^\ssr{II}_{n_1,n_2}$ and $\Upsilon^\ssr{III}_{\ell_1,\ell_2}$.

The expansion in \rf{0-SOL1_II} holds both for the
$\Upsilon_{k_1,k_2}^\ssr{II,III}(\vartheta^\circ,\varphi^\circ)$
functions as well as for the $\Psi_{k_1,k_2}^\ssr{II,III}(x,y)$ 
functions on the disk. Although \rf{U-CC} is an explicit formula 
for the II--III interbasis expansion coefficients 
$U_{\ell_1, \ell_2}^{n_1, n_2}$, we consider 
worthwhile to pursue alternative closed expressions that
will turn out to involve the Racah discrete polynomials,
which occupy the highest rung in the Askey-Wilson classification
\cite{Koekoek:2010}. Also, we will suggest that Wigner $6j$ 
recoupling coefficients may express this interbasis expansion.


\subsection{Considerations on parities}   

Since both the Legendre and Gegenbauer polynomials
have definite parities, so do the new Zernike solutions 
$\Psi^\ssr{II}_{n_1,n_2}(x,y)$ and 
$\Psi^\ssr{III}_{\ell_1,\ell_2}(x,y)$ in \rf{Psi-II}
and \rf{Psi-III}. These are 
\bea
	\Psi^\ssr{II}_{n_1,n_2}(x,y)
	&=& (-1)^{n_2}\Psi^\ssr{II}_{n_1,n_2}(-x,y)
	= (-1)^{n_1}\Psi^\ssr{II}_{n_1,n_2}(x,-y),\lab{parittyx}\\[5pt]
	\Psi^\ssr{III}_{\ell_1,\ell_2}(x,y)
	&=&\, (-1)^{\ell_1}\Psi^\ssr{III}_{\ell_1,\ell_2}(-x,y)\,
	\,=\, (-1)^{\ell_2}\Psi^\ssr{III}_{\ell_1,\ell_2}(x,-y),
		\lab{parittyy}
\eea
where we notice that the changes of sign in the 
1- and 2- quantum numbers are intertwined.

The parity must be the same on both sides of 
the expansion \rf{0-SOL1_II}, so with the aid
of \rf{COOR-01} we separate the sum into even
and odd parts, writing $\sum^{n}_{n_2=0}
=\sum_{n_2\ \rm even} + \sum_{n_2\ \rm odd}$.
Under the transformation $x\mapsto{-x}$, 
\rf{parittyx}--\rf{parittyy} turn \rf{0-SOL1_II} into
\be
	(-1)^{\ell_1} \Upsilon^\ssr{III}_{\ell_1,\ell_2} (\vartheta''\!, \varphi'')
		 = \!\!\!\sum_{n_2\ \rm even }\!\!\!\! U_{\ell_1, \ell_2}^{n_1, n_2}
		\Upsilon^\ssr{II}_{n_1,n_2} (\vartheta'\!, \varphi')
	- \!\!\!\sum_{n_2\ \rm odd}\!\!\!\! U_{\ell_1, \ell_2}^{n_1,n_2}
	\Upsilon^\ssr{II}_{n_1,n_2} (\vartheta'\!, \varphi'),
		 \lab{0-SOL1_II-new-1}
\ee
which compared with the original \rf{0-SOL1_II} imply 
that, when $\ell_1$ is odd or even, the coefficients 
$U_{\ell_1, \ell_2}^{n_1,n_2}$ of either the even or the odd 
part of the sum are zero,
\be
   U^{n_1,2p_2}_{2q_1+1,\ell_2}=0, \qquad
   U^{n_1,2p_2+1}_{2q_1,\ell_2}=0, \lab{Uzero1}
\ee
where we have written odd $\ell_1=2q_1+1$ and even 
$\ell_1=2q_1$, as well as even $n_2=2p_2$ and odd 
$n_2=2p_2+1$, for integer $q_1,\,p_2$.

The summation over $n_2$ in \rf{0-SOL1_II} can be turned
into a summation over $n_1=n-n_2$ with the same division 
into even and odd terms, and considered under the transformation
$y\mapsto -y$, yielding 
\be
	(-1)^{\ell_2} \Upsilon^\ssr{III}_{\ell_1,\ell_2} (\vartheta''\!, \varphi'')
		 = \!\!\!\sum_{n_1\ \rm even }\!\!\!\! U_{\ell_1, \ell_2}^{n_1, n_2}
		\Upsilon^\ssr{II}_{n_1,n_2} (\vartheta'\!, \varphi')
	- \!\!\!\sum_{n_1\ \rm odd}\!\!\!\! U_{\ell_1, \ell_2}^{n_1,n_2}
	\Upsilon^\ssr{II}_{n_1,n_2} (\vartheta'\!, \varphi').
		 \lab{0-SOL1_II-new-2}
\ee
Again comparing with the original \rf{0-SOL1_II}, we
conclude that when $\ell_1$ is odd or even, then
\be
   U^{2p_1,n_2}_{\ell_1,2q_2+1}=0, \qquad
   U^{2p_1+1,n_2}_{\ell_1,2q_2}=0, \lab{Uzero2}
\ee
for integer $p_1,\,q_2$. 

From \rf{Uzero1}
and \rf{Uzero2} we reach the result that the 
coefficients $U^{a,b}_{\alpha,\beta}$ are {\it non-zero\/}
only when $(a,\beta)$ have the same parity and also
$(b,\alpha)$ have the same parity. This leaves \rf{0-SOL1_II}
broken up into {\it four\/} separate cases.
\bea
\noalign{\noindent For even $n$ states:}
	\Upsilon^\ssr{III}_{2q_1,2q_2}(\vartheta'', \varphi'')
		&=& \sum_{p_1,p_2} U_{2q_1, 2q_2}^{2p_1, 2p_2}\,
		\Upsilon^\ssr{II}_{2p_1,2p_2}(\vartheta', \varphi'),
		\lab{E-E}\\
	\Upsilon^\ssr{III}_{2q_1+1,2q_2+1}(\vartheta'', \varphi'')
		&=& \sum_{p_1,p_2} U_{2q_1+1, 2q_2+1}^{2p_1+1, 2p_2+1}\,
		\Upsilon^\ssr{II}_{2p_1+1,2p_2+1}(\vartheta', \varphi'),
		\lab{O-O}\\
\noalign{\noindent For odd $n$ states:}
	\Upsilon^\ssr{III}_{2q_1+1,2q_2}(\vartheta'', \varphi'')
		&=& \sum_{p_1,p_2} U_{2q_1+1, 2q_2}^{2p_1, 2p_2+1}\,
		\Upsilon^\ssr{II}_{2p_1,2p_2+1}(\vartheta', \varphi'),
		\lab{O-E}\\
	\Upsilon^\ssr{III}_{2q_1,2q_2+1}(\vartheta'', \varphi'')
		&=& \sum_{p_1,p_2} U_{2q_1, 2q_2+1}^{2p_1+1, 2p_2}\,
		\Upsilon^\ssr{II}_{2p_1+1,2p_2}(\vartheta', \varphi').
		\lab{E-O}
\eea
In every case, the sum of each pair of indices will add to 
the principal quantum number $n$, and the relation 
\rf{COOR-01} between the angles will hold. In the following 
two subsections we analyse separately the two cases presented by 
even $n_2$ in \rf{E-E} and \rf{E-O}, and by odd $n_2$ 
in \rf{O-O} and \rf{O-E}, following routes parallel to 
that used in the previous section for the I--II interbasis 
coefficients.

\subsection{Interbasis coefficients $U_{\ell_1, \ell_2}^{n_1, n_2}$ for even $n_2$}  
		\label{sec:4.2}

We consider first the interbasis expansion
\rf{0-SOL1_II} for even $n_2$. The
coefficients $U_{\ell_1, \ell_2}^{n_1, n_2}$ can
be calculated again as we did following
\rf{angles-II-I} for $\vartheta' =
\pi/2-\varepsilon$ on the common rim of
the disk and half-sphere. As before, in
the limit $\varepsilon\to0$ on both sides of
\rf{0-SOL1_II-new-2}, with $\cos\vartheta''
= \cos\varphi'$, $\cos\varphi'' \approx 1$, and
$\sin\varphi''\approx \cos\vartheta'/\sin\varphi'$
in \rf{Syst-II} and \rf{Syst-III}, where $x=\xi_1=0$.
Using the expressions for Legendre and Gegenbauer
polynomials in \rf{Psi-II} and \rf{Psi-III} for
even $\ell_1=2q_1$ and $n_2=2p_2$,
with $q_1,p_2$ non-negative integers,
\be
	P_{2q_1}(0) = \frac{(-1)^{q_1}}{2^{2q_1}}
		\frac{(2q_1)!}{(q_1!)^2},  \qquad
	C_{2p_2}^{n_1 + 1}(0) = (-1)^{p_2}
		\frac{(n_1+p_2)!}{n_1!\, p_2!}, \lab{PCpar}
\ee
then multiplying \rf{0-SOL1_II} by
$P_{n_1'}(x)\,\dd x$, integrating over the region
$\cos\varphi'=x\in[-1,1]$, and taking into account
the orthogonality and square norm of
Legendre polynomials,
$\int_{-1}^1 \dd x\, [P_{\ell}(x)]^2
= 1/(\ell+\onehalf)$, we obtain after integration
\bea
	U_{2q_1, \ell_2}^{n_1, 2p_2} &=&
		A_{q_1, \ell_2}^{n_1, p_2} \, \int_{-1}^1 \dd x\,(1-x^2)^{q_1}
		\, C_{\ell_2}^{2q_1+1} (x)\, P_{n_1}(x), \lab{INT-01} \\
\noalign{\noindent with the coefficient}
	A_{q_1, \ell_2}^{n_1, p_2} &=&
		\frac{(-1)^{q_1-p_2}}{2^{n_1}}
			\frac{[(2q_1)!]^2 \, p_2!}{(q_1!)^2\,(n_1+ p_2)!} \nonumber\\[3mm]
		&&{}\times
	\sqrt{(2q_1 + \onehalf)(n_1+\onehalf)
		\frac{\ell_2! \,(2n_1 + 2p_2 +1)!}{(2p_2)!\,(4q_1 + \ell_2 + 1)!}}. \lab{INT-02}
\eea

To solve the integral \rf{INT-01} we use \rf{APP-05} and \rf{APP-07} 
from Appendix A, for
	$$\begin{array}{ll}
       \ell_1= 2q_1,& n_1= 2p_1,\\
       \ell_2= 2q_2,& n_2=2p_2,\end{array}
	\quad \hbox{and}\quad 2q_1+2q_2 = n = 2p_1 + 2p_2,$$
to obtain the even-even coefficients \rf{E-E} written in 
terms of $_4F_3$ hypergeometric polynomials as
\bea
	U_{2q_1, 2q_2}^{2p_1, 2p_2} &=&
	\frac{(-1)^{q_2}}{2^{2p_1}} \frac{p_2!}{p_1!} \frac{[(2q_1)!]^2}{(4q_1+1)!}
	\frac{\sqrt{(2p_1+\onehalf)(2q_1+\onehalf)}}{(2p_1+p_2)!}\nonumber\\[5pt]
		&&{}\times
	\frac{\Gamma(p_1 + \frac12)}{(q_1 - p_1)!\,\Gamma(q_1+ p_1 + \frac32)}
	\sqrt{\frac{(4p_1+ 2p_2 + 1)!\,(4q_1+ 2q_2 + 1)!}{(2p_2)!\,(2q_2)!}}	
		\nonumber\\[5pt] &&{}\times
		{_4F_3}\bigg(\begin{array}{c} - q_2, \quad 2q_1+q_2+1, \quad q_1+1, \quad q_1 + 1 \\
			2q_1+\frac32,\quad q_1 + p_1+ \frac32, \quad q_1-p_1 + 1
				\end{array}\bigg|\,1\bigg)\lab{INT-03}.
\eea

On the other hand, for
	$$\begin{array}{ll}
       \ell_1=2q_1,& n_1 = 2p_1+1,\\
       \ell_2 = 2q_2+1,& n_2=2p_2,\end{array}
       \quad \hbox{and}\quad 2q_1+2q_2 + 1 =n= 2p_1 + 2p_2 + 1,$$
we use the same formulas from the Appendix to 
write \rf{E-O} as
\bea
	U_{2q_1, 2q_2+1}^{2p_1+1, 2p_2} &=&
	\frac{(-1)^{q_2}}{2^{2p_1+1}} \frac{p_2!}{p_1!}
		\frac{[(2q_1)!]^2}{(2p_1+p_2+1)!\,(4q_1+1)!}
	\sqrt{\frac{(2p_1+\frac32)(2q_1+\frac12)}{(2p_2)!\,(2q_2+1)!}} \nonumber\\[5pt]
		&&{}\times
	\frac{\Gamma(p_1 + \frac32)\,\sqrt{(4p_1+ 2p_2 + 3)!\,(4q_1+ 2q_2 + 2)!}
		}{\Gamma(q_1 - p_1+1)\,\Gamma(q_1+ p_1 + \frac52)} \nonumber\\[5pt]
	&&{}\times
	{_4F_3}\bigg(\begin{array}{c} - q_2, \quad 2q_1+q_2+2, \quad q_1+1, \quad q_1 + 1\\
	2q_1+\frac32,\quad q_1 + p_1 + \frac52, \quad q_1 -p_1 +1  \end{array}
		\bigg|\,1\bigg). \lab{INT-04}
\eea


\subsection{Interbasis coefficients $U_{\ell_1, \ell_2}^{n_1, n_2}$ for odd $n_2$}

We now consider the interbasis expansion coefficients 
$U_{\ell_1, \ell_2}^{n_1, n_2}$ in \rf{0-SOL1_II} for odd 
$n_2=2p_2+1$. We divide both sides of this expansion 
by $\cos\vartheta'$ and again take the limit $\vartheta' \to 
\frac12{\pi}-\varepsilon$ for $\varepsilon\to0$. 
We require the following limit expressions
for Legendre and Gegenbauer polynomials when the quantum
numbers $\ell_1=2q_1+1$ and $n_2=2p_2+1$ are odd,
\bea
	\left.\frac{P_{2q_1+1}(\sin\varphi'')}{\cos\vartheta'}
		\right|_{\vartheta' \to \frac12{\pi}} \!\!\!&=&
		\frac{(-1)^{q_1} (\frac32)_{q_1}}{q_1!\, \sin\varphi' }, \lab{2-INT-01}\\
	\left. \frac{C_{2p_2+1}^{n_1 + 1}(\cos\vartheta')}{\cos\vartheta'}
		\right|_{\vartheta' \to \frac12{\pi}} \!\!\! &=& 2 (-1)^{p_2} \frac{(n_1+p_2+1)!}{n_1!\, p_2!}.
		\lab{2-INT-01-bis}
\eea
Using the orthogonality relation for Legendre polynomials,
and by analogy with the previous case of even indices, we obtain
\bea
	U_{2q_1+1, \ell_2}^{n_1, 2p_2+1} &=&  B_{q_1, \ell_2}^{n_1, p_2} \, \int_{-1}^1\dd x\, (1-x^2)^{q_1}
		\, C_{\ell_2}^{2q_1+2}(x)\, P_{n_1}(x) ,  \lab{2-INT-02}\\
\noalign{\noindent with}
	B_{q_1, \ell_2}^{n_1, p_2}
	&=& \frac{(-1)^{q_1-p_2}}{2^{n_1} }  \frac{p_2!}{(n_1+p_2+1)!}
		\sqrt{\frac{(2p_2+2n_1+2)!\, \ell_2!}{(\ell_2+ 4q_1+3)!\,(2p_2 + 1)!}} \nonumber\\
	&&{}\times \sqrt{(n_1 + \onehalf)(2q_1+ \tsty32)}
	\left(\frac{(2q_1+1)!}{q_1!}\right)^2. \lab{2-INT-03}
\eea

As before, we must consider separately
two cases: when $\ell_2$, $n_1$ are both even or both odd.
Using again formulas \rf{APP-05} and \rf{APP-07} from 
Appendix A we obtain, for $\ell_2=2q_2$ and $n_1 = 2p_1$,
thus $n=2p_1 + 2p_2 +1=2q_1 +2q_2 +1$,
\be
	\begin{array}{l} \displaystyle
	{\hskip-15pt}U_{2q_1+1, 2q_2}^{2p_1, 2p_2+1}
		=  \frac{(-1)^{q_2}}{2^{2p_1}} \frac{p_2!\,[(2q_1+1)!]^2}{p_1!\, (2p_1 + p_2+1)!}
		\sqrt{\frac{(2p_1+\tsty12)(2q_1+\tsty32)}{(2p_2 + 1)!\, (2q_2)!}} \\[10pt]
	\displaystyle{}\times \frac{\Gamma(p_1+\frac12)}{\Gamma(4q_1+4)}\,
	\frac{\sqrt{(2p_1+n+1)!\,(2q_1+n+2)!}
		}{\Gamma(q_1+p_1+\tsty32)\,\Gamma(q_1-p_1+1)}
	\\[10pt]
	{}\times {_4F_3}\bigg(\begin{array}{c} - q_2, \quad q_2+2q_1+2, \quad  q_1+1, \quad  q_1 + 1\\
		2q_1+\frac52,\quad  q_1 + p_1 + \frac32, \quad  q_1 -p_1 +1 \end{array}\bigg|\,1\bigg),
	\end{array}\lab{2-INT-04}
\ee
while for $\ell_2=2q_2+1$, $n_1 = 2p_1+1$, thus $n= 2p_1 + 2p_2 +2 = 2q_1 +2q_2 +2$,
\be
	\begin{array}{l}\displaystyle
	{\hskip-15pt}U_{2q_1+1, 2q_2+1}^{2p_1+1, 2p_2+1} =  \frac{(-1)^{q_2}}{2^{2p_1+1}}
	\frac{p_2!\,[(2q_1+1)!]^2}{p_1!\, (2p_1 + p_2+2)!} \sqrt{\frac{(2p_1+\tsty32)
		(2q_1+\tsty32)}{(2p_2 + 1)!\, (2q_2 + 1)!}}\\[10pt] \displaystyle
	{}\times \frac{\Gamma(p_1+\tsty32)
		}{\Gamma(4q_1+4)}\,\frac{\sqrt{(2p_1+n+2)!\,(2q_1+n+2)!}
		}{\Gamma(q_1+p_1+\tsty52)\,\Gamma(q_1-p_1+1)}\\[10pt]
	{}\times {_4F_3} \bigg(\begin{array}{c} - q_2, \quad  q_2+2q_1+3, \quad  q_1+1, \quad  q_1 + 1,\\
	2q_1+\frac52,\quad  q_1 + p_1 + \frac52, \quad  q_1 -p_1 +1  \end{array}\bigg|\,1\bigg).
	\end{array} \lab{2-INT-05}
\ee
The four cases of nonzero II-III interbasis coefficients 
$U_{\ell_1, \ell_2}^{n_1, n_2}$ that appear in Eqs.\ \rf{INT-03} and
\rf{INT-04} for even $n_2$, and in Eqs.\ \rf{2-INT-04} 
and \rf{2-INT-05} for odd $n_2$, have  thus been given in terms
of $_4F_3(\cdots|1)$ hypergeometric polynomials. We now proceed
to express them in terms of Racah polynomials.


\subsection{Coefficients $U_{\ell_1, \ell_2}^{n_1, n_2}$ as Racah polynomials}

Recall that in the case of the I-II interbasis 
expansion in Sect.\ \ref{sec:three}, the coefficients 
$W^{n, m }_{n_1, n_2}$, depending on three effective 
parameters as is evident in their Clebsch-Gordan
form \rf{COEF2}, were written in terms of 
$_3F_2(\cdots|1)$'s and as Hahn polynomials, the latter two having 
five avaliable parameters. Now we have written the four
distinct nonzero sets of II-III interbasis coefficients
$U_{\ell_1, \ell_2}^{n_1, n_2}$, which also depend on three
effective parameters, in terms of $_4F_3(\cdots|1)$'s
that in principle can provide {\it seven\/} 
available parameters. These have a special form though:
First let us recall that when the parameters in a 
hypergeometric series
$$
	_{k+1}F_k(a_1,a_2,\ldots,a_{k+1}; b_1,\ldots,b_k;z)
$$	
\vskip-10pt
\noindent are such that
\vskip-10pt
\be
	a_1 + a_2 + \cdots + a_{k+1} + 1 \,=\, b_1 + b_2 + \cdots + b_k, 
		\lab{Saal}
\ee
the series is called {\it balanced}, or {\it Saalsch{\"u}tzian\/}
(see \cite[p.\ 188]{BE1}). 
It is not difficult to verify that the ${_4F_3}$'s in
\rf{INT-03}, \rf{INT-04}, \rf{2-INT-04} and \rf{2-INT-05} 
satisfy this condition and could enjoy six free parameters. 

Saalsch\"ultzian hypergeometric polynomials can be expressed in terms of
Racah polynomials of degree $n$ in the variable $x$, also with 
with six effective parameters $\alpha, \beta, \gamma, \delta$, 
plus $n$ and $x$, as
\be
	R_n( \lambda\of{x}; \alpha, \beta, \gamma, \delta)
		:={_4F_3} \bigg(\begin{array}{c} - n,\ n{+}\alpha{+}\beta{+}1,
			\ -x,\  x{+}\gamma{+}\delta{+}1\\
		\alpha{+}1,\ \beta{+}\delta{+}1,\ \gamma{+}1 \end{array}
		\bigg|\,1 \bigg), \lab{Racah}
\ee
on the quadratic lattice $\lambda\of{x}
:=x(x+\gamma + \delta +1)$ of $x\in\{0,1,\ldots,N\}$. 
The range of degrees of the polynomials is 
$n\in\{0,1,2,\ldots,N\}$, where $N$ a nonnegative integer 
which can have one of three values according to whether
\cite[Eq.\ (9.2.1)]{Koekoek:2010}: 
\be
	\alpha + 1 = - N,\quad \hbox{or} \quad \beta + \delta = - N,
		\quad \hbox{or} \quad \gamma +1= -N.  \lab{rangesn}
\ee
In {\it each\/} of the ranges \rf{rangesn}, a set of Racah 
polynomials is orthogonal over the points
in the quadratic lattice $\lambda\of{x}$, with 
weight functions $\rho\of{x}$ and norms $d_n$, as was the
case of the Hahn polynomials in \rf{orthoHahn}. Their use will provide more 
compact formulas below. But first we must transform the
$_4F_3(\cdots|1)$ hypergeometrics to their canonical form
\rf{Racah}; this is done in Appendix B, and results in
the following forms for  $_4F_3(\cdots|1)$'s expressible
with Racah polynomials. 

\subsubsection{The coefficients $U_{2q_1, 2q_2}^{2p_1, 2p_2}$}

Substituting now \rf{Finstep} into \rf{INT-03}, we obtain the interbasis 
coefficients for the even-even coefficients $U_{2q_1, 2q_2}^{2p_1, 2p_2}$ 
in \rf{E-E}, where the $_4F_3$ parameters can be readily 
compared with their `Racah form' in  \rf{Racah}. Defining here the number 
$N:= p_1+p_2= q_1+q_2$ for all subsequent expressions, we write
\bea
	U_{2q_1, 2q_2}^{2p_1, 2p_2}
	&=&(-1)^{p_1+ q_2} 4^{q_1+p_1} \frac{(q_1+N)!\,(p_1+N)!}{q_2!\,p_2!} \nonumber \\
	 &&{\quad}\times \sqrt{\frac{(4q_1+1)(4p_1+1)\,(2q_2)!\,(2p_2)!
		}{(2q_1+ 2N + 1)!\,(2p_1+ 2N + 1)!}} \lab{INT-03-F} \\
	&&{\quad}\times {_4F_3}\bigg(\begin{array}{c}
		-p_1, \quad p_1 + \frac12, \quad  q_1 + \frac12, \quad -  q_1 \\
		 N + 1, \quad  1 , \quad  - N \end{array}\bigg|\,1\bigg)\nonumber\\
		 &=&(-1)^{p_1+ q_2}\,\frac{\sqrt{\rho(p_1)}}{d_{q_1}}\,
		R_{q_1}\Big( \lambda(p_1); \alpha, \beta, \gamma, \delta\Big) \\
		&=& (-1)^{p_1+ q_2}\,\frac{\sqrt{\rho(q_1)}}{d_{p_1}}\,
			R_{p_1}\Big( \lambda(q_1); \alpha, \beta, \gamma, \delta\Big).
				 \lab{42'}  
\eea
The identification with the parameters in \rf{Racah} has 
$N=\onehalf n$ (because here $n=n_1+n_2=2p_1+2p_2
=\ell_1+\ell_2=2q_1+2q_2$), the quadratic lattice 
$\lambda(x)=x(x+\onehalf)$, the parameters $\alpha= N$, 
$\beta = - \delta = - (N{+}\onehalf)$, and
$ \gamma = - (N{+} 1)$, and the weight and norm factors
\bea	
	\rho(p_1) &:=&4^{2p_1}\frac{(2N+1)(4p_1+1)\,(2p_2)!\,
		[(p_1+N)!]^2}{(2p_1+2N+1)!\,(p_2!)^2}\,, \nonumber\\ 
	d_{q_1} &:=& \displaystyle \frac{q_2!}{4^{q_1}(q_1+N)!}\,
		\sqrt{\frac{(2N+1)(2k_1+2N+1)!}{(4q_1+1)(2q_2)!}}\,\,.
		\nonumber
\eea
The $_4F_3$ hypergeometric in \rf{INT-03-F} represents 
a particular family of self-dual Racah polynomials
$R_n(\lambda\of{x}; \alpha, \beta, \gamma, \delta)$
because its parameters are interconnected by
$\alpha + \beta = \gamma + \delta$, which
means that this $_4F_3$ can be expressed by
two equivalent Racah polynomials in the
discrete variable,
\be
	 R_{q_1}\Big( \lambda(p_1); \alpha, \beta, \gamma, \delta\Big)=
	 R_{p_1}\Big( \lambda(q_1); \alpha, \beta, \gamma, \delta \Big),  \lab{Rac1}
\ee
over the same quadratic lattice $\lambda\of{x}=x(x+\onehalf)$ 
and the same parameters $\alpha$, $\beta$, $\gamma$ and $\delta$.

\subsubsection{The coefficients $U_{2q_1, 2q_2+1}^{2p_1+1, 2p_2}$}

For the coefficients $U_{2q_1, 2q_2+1}^{2p_1+1, 2p_2}$ 
in \rf{O-E}, the $_4F_3$ hypergeometric in \rf{INT-04} can
be again transformed as in Appendix B to the canonical form
\rf{Racah}, which simplifies to
\bea
	U_{2q_1, 2q_2+1}^{2p_1+1, 2p_2}
	&=&(-1)^{p_1+ q_2} 2^{2q_1+ 2p_1+1} \frac{(q_1+N+1)!\, (p_1+N+1)!}{(N+1)\,q_2!\,p_2!}\nonumber\\
	 &&{\quad}\times\sqrt{
		\frac{(4q_1+1)\,(4p_1+3)\,(2q_2 + 1)!\,(2p_2)!}{(2q_1+ 2N + 2)!\,(2p_1+ 2N + 3)!}}\lab{INT-04-F}\\
	 &&{\quad}\times{_4F_3}\bigg(\begin{array}{c}
			 - p_1, \quad p_1 + \frac32, \quad  q_1 + \frac12, \quad -q_1 \\
		 N + 2, \quad  1 , \quad  -  N \end{array}\bigg|\,1\bigg)\nonumber\\
	&=&(-1)^{p_1+ q_2}\,\frac{\sqrt{\rho_1(q_1)}}{d_{p_1}^{(1)}}\,
		R_{p_1}\Big( \lambda(q_1); \alpha, \beta, \gamma, \delta\Big)\lab{RR1}\\
		&=& (-1)^{p_1+ q_2}\,\frac{\sqrt{\rho_2(p_1)}}{d_{q_1}^{(2)}}\,
		R_{q_1}\Big( \mu(p_1); \alpha, \beta-1, \gamma, \delta+1\Big).\lab{RR2}
\eea
Here $N:= p_1 + p_2 = q_1 + q_2=  \onehalf(n-1)$, 
the parameters are $\alpha=-\gamma= N+1$ and 
$\beta = -\delta = -(N + \onehalf)$, 
in the two last expressions the quadratic lattices are 
$\lambda(x)=x(x+\onehalf)$ and $\mu(x)=x(x+\frac32)$,
and the weight and norm factors are
\bea
	\rho_1(q_1) &=& 2^{4q_1+1}\frac{(4q_1+1)(2q_2+1)![(q_1+N + 1)!]^2
		}{(N+1)(2q_1+2N+2)!(q_2!)^2}\,,\nonumber \\ [2mm]
	d_{p_1}^{(1)} &=& \frac{p_2!}{4^{p_2}(p_1+N+1)!}\,
		\sqrt{\frac{(N+1)(2p_1+2N+3)!}{2(4p_1+3)(2p_2)!}}\,,\nonumber \\ [2mm]
	\rho_2(p_1) &=& 2^{4p_1+1}\frac{(2N+1)(2N+3)(4p_1+3)(2p_2)![(p_1+N+1)!]^2
		}{3(N+1)(2p_1+2N+3)!(p_2!)^2}\,,\nonumber\\ [2mm]
	d_{q_1}^{(2)} &=& \frac{q_2!}{4^{q_1}(q_1+N+1)!}\,
		\sqrt{\frac{(N+1)(2N+1)(2N+3)(2k_1+2N+2)!}{6(4q_1+1)(2q_2+1)!}}\,. \nonumber
\eea

\subsubsection{The coefficients $U_{2q_1+1, 2q_2}^{2p_1, 2p_2+1}$}

	Regarding the interbasis coefficients for 
odd $n_2$ in \rf{2-INT-04}, and performing the $_4F_3$ transformations
of Appendix B, we obtain
\bea
	U_{2q_1+1, 2q_2}^{2p_1, 2p_2+1} &=&  {(-1)^{q_2+p_1}} 2^{2q_1+2p_1+1}
	\frac{(N+q_1+1)!\, (N+p_1+1)!}{(N+1)\, q_2!\, p_2!} \nonumber \\
	  &&{}\times
	\sqrt{\frac{(4p_1+1) (4q_1+3)\, (2p_2+1)!\, (2q_2)! }{(2N+2p_1+2)!\, (2N+2q_1+3)!}}\lab{2-INT-042}\\
	  &&{}\times
		{_4F_3}\bigg(\begin{array}{c} -p_1, \quad  p_1 +\frac12,\quad q_1+\frac32,\quad -q_1 \\
		 N + 2,\quad   1 , \quad   -  N \end{array}\bigg|\,1\bigg)\nonumber\\ 
	 &=&(-1)^{p_1+ k_2}\,\frac{\sqrt{\rho_1(p_1)}}{d_{q_1}^{(1)}}\,
		R_{q_1}\Big( \lambda(p_1); \alpha, \beta, \gamma, \delta\Big)\lab{co} \\ 
	 &=&(-1)^{p_1+ q_2}\,\frac{\sqrt{\rho_2(q_1)}}{d_{p_1}^{(2)}}\,
		R_{p_1}\Big( \mu(q_1); \alpha, \beta-1, \gamma, \delta+1\Big).\lab{58'}
\eea
Here again $N:= p_1 + p_2=q_1 + q_2=  \onehalf(n-1)$, 
but the parameters are now $\alpha=-\gamma = N{+}1$ and  
$\beta = - \delta = - (N{+}\onehalf)$,
the quadratic lattices are  $\lambda(x)=x(x+\tsty12)$ 
and $\mu(x)=x(x+\tsty32)$, and the weight and norm factors are 
\bea
	\rho_1(p_1) &=& 2^{4p_1+1}\frac{(4p_1+1)\,(2p_2+1)!\,[(p_1+N + 1)!]^2
		}{(N+1)\,(2p_1+2N+2)!\,(p_2!)^2}\,,\nonumber\\ 
	d_{q_1}^{(1)} &=& \frac{q_2!}{4^{q_1}\,(q_1+N+1)!}
		\sqrt{\frac{(N+1)\,(2q_1+2N+3)!}{2(4q_1+3)\,(2q_2)!}}\,,\nonumber \\ 
	\rho_2(q_1) &=& 2^{4q_1{+}1}\frac{(2N{+}1)(2N{+}3)(4q_1{+}3)\,(2q_2)!\,[(q_1{+}N{+}1)!]^2
			}{3(N{+}1)\,(2q_1{+}2N{+}3)!\,(q_2!)^2}\,, \nonumber\\ 
	d_{p_1}^{(2)} &=& \frac{p_2!}{4^{p_1}\,(p_1{+}N{+}1)!}
		\sqrt{\frac{(N{+}1)(2N{+}1)(2N{+}3)\,(2p_1{+}2N{+}2)!}{6(4p_1{+}1)\,(2p_2{+}1)!}}. \nonumber
\eea

\subsubsection{The coefficients $U_{2q_1+1, 2q_2+1}^{2p_1+1, 2p_2+1}$}

Finally, the odd-odd interbasis coefficients in \rf{2-INT-05} 
can be brought in terms of Racah polynomials as
\bea
	U_{2q_1+1, 2q_2+1}^{2p_1+1, 2p_2+1} &=&  {(-1)^{q_2+p_1}} 2^{2q_1+2p_1+2}
	\frac{(N+q_1+2)!\,(N+p_1+2)!}{(N+1)(N+2)\, q_2!\, p_2!}\nonumber \\
	&&{}\times
	\sqrt{\frac{(4p_1+3) (4q_1+3)\, (2p_2+1)! \,(2q_2+1)! }{(2N+2p_1+4)!\, (2N+2q_1+4)!}}
	\lab{2-INT-052}	\\ &&{}\times
	{_4F_3}\bigg(\begin{array}{c} -p_1,\quad p_1+\frac32,\quad q_1+\frac32,\quad -q_1 \\
	 N + 3,\quad   1 , \quad   -  N \end{array}\bigg|\,1\bigg) \nonumber\\
	&=&(-1)^{p_1+ q_2}\,\frac{\sqrt{\rho(p_1)}}{d_{q_1}}\,
		R_{q_1}\Big( \mu(p_1); \alpha, \beta, \gamma, \delta\Big)\lab{59'} \\ 
	&=&(-1)^{p_1+ q_2}\,\frac{\sqrt{\rho(q_1)}}{d_{p_1}}\,
		R_{p_1}\Big( \mu(q_1); \alpha, \beta, \gamma, \delta\Big). \lab{59c}
\eea
Here $N:= p_1 + p_2=q_1 + q_2=  \onehalf n-1$, the parameters are 
$\alpha=1{-}\gamma= N{+}2$, $\beta = -\delta = -(N{+}\tsty32)$,
the lattice is  $ \mu(x)=x(x+\tsty32)$, and the weight and norm factors are
\bea
	\rho(q_1) &=& 4^{2q_1+1}\frac{(2N+3)(4q_1+3)\,(2q_2+1)!\,[(q_1+N+2)!]^2
		}{3(N+1)(N+2)\,(2q_1+2N+4)!\,(q_2!)^2}\,,\nonumber\\
	d_{p_1} &=& \frac{p_2!}{2^{2p_1+1}\,(p_1+N+2)!}
		\sqrt{\frac{(N+1)(N+2)(2N+3)\,(2p_1+2N+4)!}{3(4p_1+3)\,(2p_2+1)!}}.\nonumber
\eea
Our final remark is to point out that the 
expressions for all II-III interbasis coefficients
$U_{\ell_1, \ell_2}^{n_1, n_2}$ are equivalent to
the summation formula \rf{U-CC2} of the product of
two special Clebsch-Gordan coefficients.


\subsection{Relation with the Wigner $6j$ symbols}

The coefficients that bridge two distinct coupling 
orders between three spins to the same total spin
are known as Wigner $6j$ symbols. They contain
six spin parameters: $\ell_1,\,\ell_2,\,\ell_3$, 
their couplings to $\ell_{12},\,\ell_{23}$
and the total $\ell$, and have a host of symmetry 
relations that can be seen in the literature 
\cite{VilKl}.  These $6j$ coefficients can be 
expressed in terms of balanced $_4F_3(\cdots|\,1)$ 
functions (see for example Refs.\ \cite{V,VilKl}), 
so they can be also written in terms of Racah polynomials 
through \rf{Racah}, having the same number of parameters, 
times a lengthy factor containing factorials and Kronecker 
triangle functions (1 when they couple properly, 
0 if not).

In Subsec.\ \ref{sec:Clebsch} we wrote the I-II
the interbasis coefficients $W_{n_1, n_2}^{n,m}$, 
containing three effective labels, in terms of {\it proper\/} 
Clebsch-Gordan coefficients, i.e., whose three 
spin indices and their projections are integer 
or half-integer, form a triangle and vanish
when not. On the other hand, their expression as 
Hahn polynomial functions in \rf{COEF12}, allows
for analytic continuation in all parameters. 

Here we find that it is {\it not\/} always 
the case that the II-III interbasis coefficients 
$U_{\ell_1, \ell_2}^{n_1, n_2}$ correspond to 
{\it proper\/} $6j$ coefficients. Yet their 
relation is sufficiently close to merit attention. 
We thus proceed to examine the known equivalence
between Wigner $6j$ symbols and balanced 
$_4F_3(\cdots|\,1)$ hypergeometric functions, 
which is \cite[\S 8.4.4]{VilKl}
\bea 
	&&{}\hskip-15pt\left\{\begin{array}{ccc}\ell_1 & \ell_2 & \ell_{12} \\ 
			\ell_3 & \ell & \ell_{23}\end{array}\right\} \lab{6J-4F3}\\
	&&{}= c_0\, {_4F_3}\bigg(\begin{array}{c}
		\ell_1{-}\ell_2{-}\ell_{12}, \ \ell_3{-}\ell_2{-}\ell_{23}, 
		\ -\ell_1{-}\ell_2{-}\ell_{12}{-}1, \ -\ell_2{-}\ell_3{-}\ell_{23}{-}1 \\
 		-2 \ell_2,\ \ell{-}\ell_2{-}\ell_{12}{-}\ell_{23}, 
 			\ -\ell_2{-}\ell_{12}{-}\ell{-}\ell_{23}{-}1
				\end{array}\bigg|\,1\bigg).\nonumber
\eea

We shall consider only the case of even-even 
coefficients $U_{2q_1, 2q_2}^{2p_1, 2p_2}$ in
\rf{INT-03}, which is sufficiently illustrative
for our purpose, concentrate on the $_4F_3$ 
functions, and avoid long and distracting pre-factors 
with the notation $c_i$ for those that are not essential 
to our present endeavour. Comparing the components 
of the ${_4F_3}$ in \rf{INT-03} with those in \rf{6J-4F3},
\be 
	\begin{array}{rclrcl}
		\ell_1 {-} \ell_2 {-} \ell_{12} &=& -q_2,
			& \ell_3 {-} \ell_2 {-} \ell_{23} &=& 2q_1 {+} q_2 {+}1,\\
		- \ell_1 {-} \ell_2 {-} \ell_{12} {-} 1 &=& q_1 {+} 1,
			&  -\ell_2 {-} \ell_3 {-} \ell_{23} {-} 1 &=& q_1{+}1; \\
		-2 \ell_2 &=& q_1 {-} p_1 {+} 1, 
			&   \ell {-} \ell_2 {-} \ell_{12} {-} \ell_{23} &=& 2q_1 {+}\frac32,\\
 	 & & &\hskip-20pt -\ell_2 {-} \ell_{12} {-} \ell {-} \ell_{23} {-} 1 
 	 	&=& q_1 {+} p_1 {+} \frac32, \end{array}\nonumber
\ee
and solving this system of 6 simultaneous equations, we obtain
\be
	U_{2q_1, 2q_2}^{2p_1, 2p_2} = c_1 \left\{\begin{array}{ccc}
		\ell_1 & \ell_2 & \ell_{12} \\ \ell_3 & \ell & \ell_{23}
			\end{array}\right\},\lab{6j}
\ee
with
$$
	\begin{array}{lll}
 \ell_1 = -\onehalf{N} - 1, 
 	& \ell_2 =\onehalf(p_1-q_1-1), 
 	& \ell_{12} = \onehalf(q_2 - p_1 - 1);  \\
 \ell_3 =  \onehalf(N-1), 
 	& \,\ell\, = \onehalf(q_1 - p_1 - 1), 
 	& \,\ell_{23} = -1 - q_1 - \onehalf(p_1+q_2), 
 	\end{array} 
$$
where $N:= q_1+q_2=p_1+p_2=\onehalf n$ is a nonnegative 
integer in every case, so all parameters of the $6j$ 
symbol in \rf{6j} are integer or half-integer. 

That some of these parameters
appear with negative signs is not a problem,
because with the help of a `mirror' 
transformations \cite[\S 9.4]{V} one can 
invert some $\ell_i \to -\ell_i{-}1$, 
with a sign on the $6j$ coefficient, 
and use it for example in 
$\ell_1 = -\onehalf{N}{-}1 \to \onehalf{N}$.
What cannot be ascertained is that all triangle
relations (e.g.\ $|\ell_1{-}\ell_2|\le\ell_{12}
\le\ell_1{+}\ell_2$, etc.) are fulfilled.
On the other hand, the II-III interbasis 
coefficients are well defined with Racah 
polinomials, regardless of these relations.

\section{Conclusion}   \label{sec:five}

The Zernike system \rf{Zernikeq} in its
classical and quantum realizations 
\cite{PWY,PSWY} is superintegrable and
harbours several remarkable geometric
and spectral properties. In this paper
we highlighted its relevance to special
function theory. 

In three sets of coordinates on the
unit disk, the solutions to Eq.\ 
\rf{Zernikeq} involve Legendre, Gegenbauer
and Jacobi polynomials (and phases), as
illustrated in Fig.\ \ref{fig:tres-relaciones},
characterized each by two quantum numbers. 
Between them, the I-II and I-III relations
are given by Hahn polynomials, and II-III
by Racah polynomials in three discrete 
parameters. All relations can be expressed
also with Clebsch-Gordan coefficients, whose
geometric interpretation still eludes us,
while the role of $6j$ coefficients has only
been suggested.

Finally, we underline the fact that
the original Zernike polynomials have a great
practical importance in phase-contrast microscopy
and in the correction of wavefronts in circular 
pupils. Recent work \cite{Esp2,Esp3,Esp4,Esp1} has
extended this technique to pupils of essentially
arbitrary shape through diffeomorphisms that
conserve their basic properties. This has been
applied to describe wavefronts in sectorial, 
annular and polygonal-shaped pupils, the latter
specifically tailored to the hexagonal components
of large astronomial mirrors. As remarked in 
\cite{PWY-JOSAA}, the fact that among the members
of each horizontal-$n$ `multiplet' 
$\Psi^\ssr{II}_{0,n}$ in \rf{Psi-II} and
$\Psi^\ssr{III}_{0,n}$ in \rf{Psi-III} are
plane wave-like solutions can be of some relevance
for applications in correcting cylindrical aberrations. 

\section*{Appendix A}

To find the expressions \rf{2-INT-04}--\rf{2-INT-05},
consider the integral
\be
	J^{\mu,\lambda}_{n,m}:=
	\int_{-1}^1 \dd x\,(1-x^2)^{\mu} \, 
	C_{n}^{\lambda} (x) P_{m}(x). \lab{APP-01}
\ee
Because of the oddness of the integrand, this integral 
is nonzero only when $n$ and $m$ are both even or both
odd. We use the following expressions for the Gegenbauer 
and Legendre polynomials,
\bea
	C^{\lambda}_n (x) &=& \left\{\begin{array}{ll}
	\displaystyle \frac{(2\lambda)_{2q}}{(2q)!} \,
	{_2F_1} \bigg(\begin{array}{c} -q,\quad q + \lambda\\ \lambda + \frac12\end{array}\bigg|\, 1 - x^2 \bigg),
		&  n=2q, \\[10pt]
	\displaystyle\frac{(2\lambda)_{2q+1}}{(2q+1)!}\, x\,\,
	{_2F_1} \bigg(\begin{array}{c}- q,\quad q + \lambda + 1\\ \lambda + \frac12\end{array}\bigg|\, 1 - x^2 \bigg),
		 &  n=2q+1, \end{array}\right. \lab{APP-02} \\[5pt]
	P_m (x) &=& \left\{\begin{array}{ll}
	\displaystyle (-1)^{p} \frac{(\frac12)_p}{p!} \,
	{_2F_1} \bigg(\begin{array}{c} - p,\quad p+ \frac12\\ \frac12\end{array}\bigg|\, x^2 \bigg),
		&  m=2p, \\[10pt]
	\displaystyle (-1)^{p} \frac{(\frac32)_p}{p!} \, x \,\,
	{_2F_1} \bigg(\begin{array}{c} -p,\quad p+ \frac32\\ \frac32\end{array}\bigg|\, x^2 \bigg),
		&  m=2p+1, \end{array}\right.  \lab{APP-03}
\eea
where $(x)_n := \Gamma (x+n)/ \Gamma(x)$. Then, using
\be
	\int_{0}^1\dd y\, y^{\gamma - 1}\, (1-y)^{\rho - 1} \,
		{_2F_1}\bigg(\begin{array}{c}\alpha,\quad \beta\\ \gamma\end{array}\bigg|\, y \bigg)
		= \frac{\Gamma(\gamma)\,\Gamma(\rho)\,\Gamma(\gamma+\rho-\alpha-\beta)
			}{\Gamma(\gamma+\rho-\alpha)\,\Gamma(\gamma+\rho-\beta)}, \lab{APP-04}
\ee
with $\hbox{Re}\,\gamma > 0$, $\hbox{Re}\,\rho > 0$,
and $\hbox{Re}\,(\gamma+\rho-\alpha-\beta) > 0$, 
we must consider separately the two parity cases:

\smallskip

\noindent{\bf A.1}	When $n = 2q$ and $m = 2p$ are even, we rewrite
the integral \rf{APP-01} in form
\be
	\begin{array}{l}
	\displaystyle{\hskip-15pt} J^{\mu,\lambda}_{2q,2p}
	=\int_{-1}^1\dd x\, (1-x^2)^{\mu} \, C_{2q}^{\lambda} (x) P_{2p}(x)
		=(-1)^{p} \frac{\Gamma(2\lambda+ 2q)}{(2q)!\,\Gamma(2\lambda)}
		   \frac{\Gamma(p+\frac12)}{\Gamma(\frac12)\,p!} \\[5pt]
	\displaystyle{\quad}\times
		\sum_{s=0}^{q} \frac{(-q)_s \,(q+\lambda)_s}{(\lambda+\onehalf)_s\, s!}
	\int_{-1}^1\dd x\, (1-x^2)^{\mu+s} \,
	{_2F_1}\bigg(\begin{array}{c}-p,\quad p+\onehalf\\ 
		\onehalf\end{array}\bigg|\, x^2 \bigg).
		\end{array} \lab{APP-041}
\ee
Substituting here $x^2=y$ and using \rf{APP-04} with $\alpha = - p$, 
$\beta = p+\onehalf$, $\gamma = \onehalf$, and $\rho  = s+1+\mu$,  we obtain
\be
	\begin{array}{l}
	\displaystyle{\hskip-15pt} J^{\mu,\lambda}_{2q,2p}
	= (-1)^{p} \frac{\Gamma(2\lambda+ 2q)\,\Gamma(p + \frac12)}{\Gamma(2\lambda)\,(2q)!\,p!}
		\frac{\Gamma(\mu+1)^2}{\Gamma(\mu+ p + \frac32)\, \Gamma(\mu - p+1)} \\[5pt]
	\displaystyle{\qquad\quad}\times
	{_4F_3} \bigg(\begin{array}{c} -q, \quad  q+\lambda, \quad  \mu+1, \quad  \mu + 1 \\
		\lambda+\frac12,\quad  \mu + p + \frac32, \quad  \mu - p +1 \end{array}\bigg|\, 1 \bigg).
		\end{array} \lab{APP-05}
\ee

\smallskip

\noindent{\bf A.2} When $n = 2q+1$ and $m = 2p+1$ are odd, we have the
integral \rf{APP-01} with $\alpha = -p$, $\beta = p + 3/2$, 
$\gamma = \frac32$, and $\rho = \mu + s + 1$,
\be
	\begin{array}{l}
	\displaystyle{\hskip-15pt} J^{\mu,\lambda}_{2q+1,2p+1}
	=(-1)^{p} \frac{ \Gamma(2\lambda+ 2q + 1)\,\Gamma(p + \frac32)}{\Gamma(2\lambda)\, (2q+1)!\,p! }
	\frac{\Gamma(\mu+1)^2}{\Gamma(\mu + p + \frac52)\, \Gamma(\mu - p + 1)} \\[5pt]
	\displaystyle{\qquad\qquad}\times
	{_4F_3} \bigg(\begin{array}{c} - q, \quad  q+\lambda+1, \quad  \mu+1, \quad  \mu + 1 \\
	\lambda+\frac12,\quad  \mu + p + \frac52, \quad  \mu - p +1 \end{array}\bigg| 1 \bigg).
		\end{array} \lab{APP-07}
\ee

\section*{Appendix B}

To transform the $_4F_3(\cdots|1)$ Saalsch\"ultzian 
hypergeometric polynomials \rf{INT-03}, \rf{INT-04}, 
\rf{2-INT-04} and \rf{2-INT-05} into the canonical form  
for the Racah polynomials in \rf{Racah}, we use
the symmetry properties of terminating hypergeometric series
of the general form $_4F_3(-n,x,y,z; u,v,w;1)$
that preserve their Saalsch{\"u}tzian character. 

Two such transformation formulas come to hand: the first
is known in the literature as Whipple's formula
for termina\-ting balanced ${_4F_3}$ series
(see \cite[Eq.\ (1.7.6)]{Koekoek:2010}), namely
\be
	\begin{array}{l} \displaystyle
	{_4F_3}\bigg(\begin{array}{c} - n, \quad x, \quad  y, \quad  z\\
		u,\quad  v, \quad  w \end{array}\bigg|\,1\bigg)
			=\frac {(v-z)_n\,(u-z)_n}{(v)_n\,(u)_n} \\
	{\qquad}\times {_4F_3}\bigg(\begin{array}{c} - n, \quad w-x, \quad w-y, \quad  z\\
		1- u + z - n,\quad 1 - v + z - n, \quad  w \end{array}\bigg|\,1\bigg),
			\end{array} \lab{1TrFor}
\ee
where $(x)_n := \Gamma (x+n)/ \Gamma(x)$.
The second transformation formula that we use is,
\be
	\begin{array}{l}  \displaystyle
	{_4F_3}\bigg(\begin{array}{c} -n, \quad x, \quad y, \quad z\\
		u,\quad v, \quad w \end{array}\bigg|\,1\bigg)
			=(-1)^n\frac{(x)_n\,(y)_n\,(z)_n}{(u)_n\,(v)_n\,(w)_n} \\
	{\qquad}\times{_4F_3}\bigg(\begin{array}{c} -n,\quad 1-u-n,\quad 1-v-n,\quad  1-w-n\\
		1- x - n,\quad 1 - y - n, \quad  1-z-n \end{array}\bigg|\,1\bigg),
		\end{array} \lab{2TrFor}
\ee
which can be readily derived by reversing the order of
summation in the definition of the series.

	We can now write the expressions for the interbasis coefficients
$U_{2q_1, 2q_2}^{2p_1, 2p_2}$ in \rf{INT-03} and 
$U_{2q_1, 2q_2+1}^{2p_1+1, 2p_2}$ in \rf{INT-04} in terms of Racah polynomials
\rf{Racah} by using three successive transformations, where the first two 
are \rf{1TrFor} and \rf{2TrFor}. We start with the ${_4F_3}$
function in \rf{INT-03} and  use \rf{1TrFor} with the parameters
\be
	\begin{array}{c}
	n = q_2, \quad x = q_1 +1, \quad y =  q_1 + N + 1, \quad  z = q_1 +1, \\
	u = 2 q_1 + \tsty32, \quad v =  q_1 + p_1 + \tsty32, \quad  w = q_1 - p_1 + 1,
		\end{array} \lab{illus1}
		\ee
where $N = \onehalf n = q_1 + q_2 = p_1 + p_2$. This yields the relation
\be
	\begin{array}{l}
	{_4F_3}\bigg(\begin{array}{c} -q_2, \quad q_1 + N + 1, \quad q_1 + 1, \quad q_1 + 1\\
	 2 q_1 + \tsty32,\quad q_1 + p_1+ \tsty32, \quad q_1 - p_1 + 1 \end{array}\bigg|\,1\bigg) \\[10pt]
	\displaystyle{\quad}=\frac {(p_1+\frac12)_{q_2}\,(q_1+ \frac12)_{q_2}}{(q_1+p_1+\frac32)_{q_2}
			\,(2q_1+\frac32)_{q_2}} \\[10pt]
	{\qquad}\times {_4F_3}\bigg(\begin{array}{c} -q_2, \quad - p_1, \quad - (p_1 + N), \quad  q_1 + 1\\
	 \frac12- N,\quad \frac12- p_1 -  q_2, \quad  q_1 - p_1 + 1 \end{array}\bigg|\,1\bigg).
	  \end{array} \lab{1step}
\ee

The second step is to apply the transformation \rf{2TrFor} with the parameters
\be
	\begin{array}{c}
	n = p_1, \quad x = - (p_1 + N), \quad y =  q_1 + 1, \quad  z = - q_2 , \\
	u = \frac12 - p_1 -  q_2, \quad v =  \frac12 - N, \quad  w = q_1 - p_1 + 1,
	\end{array} \lab{illus2}
\ee
to find
\be
	\begin{array}{l}
	\displaystyle{_4F_3}\bigg(\begin{array}{c} -q_2, \quad - p_1, \quad - (p_1 + N), \quad  q_1 + 1\\
	\frac12- N,\quad \frac12- p_1 -  q_2, \quad  q_1 - p_1 + 1 \end{array}\bigg|\,1\bigg)\\[10pt]
	\displaystyle{\qquad}=(-1)^{p_1}\frac {(- p_1- N)_{p_1}\,(q_1+ 1)_{p_1}\,(- q_2)_{p_1}}
	{(\frac12 - p_1 - q_2)_{p_1}\,(\frac12 - N)_{p_1}\,(q_1 - p_1 + 1)_{p_1}} \\[10pt]
	{\qquad\quad}\times {_4F_3}\bigg(\begin{array}{c} -p_1,\quad k_2+\frac12, \quad p_2+\frac12,\quad -k_1\\
	 N + 1,\quad - ( p_1 + q_1), \quad  q_2 - p_1 + 1 \end{array}\bigg|\,1\bigg).
	 	\end{array} \lab{2step}
\ee

The third step applies to the  ${_4F_3}$ in \rf{2step}
the same transformation \rf{1TrFor}, but with the parameters
\be
	\begin{array}{c}
	n = p_1, \quad x = q_2 + \frac12, \quad y =  p_2 + \frac12, \quad  z = - q_1 ,\\
	u = - (q_1 + p_1), \quad v =  q_2 - p_1 + 1, \quad  w =  N + 1.
	\end{array} \lab{illus3}
\ee
This leads us to the form
\be
	\begin{array}{l}
	{_4F_3}\bigg(\begin{array}{c} -p_1, \quad q_2 + \frac12, \quad  p_2 + \frac12, \quad -  q_1 \\
	 N + 1,\quad - ( p_1 + q_1), \quad  q_2 - p_1 + 1 \end{array}\bigg|\,1\bigg) \\[10pt]
	\displaystyle{\qquad}= \frac {(p_2 + 1 )_{p_1}\,(- p_1)_{p_1}
		}{(q_2 - p_1  + 1)_{p_1}\,(- q_1 - p_1 )_{p_1}} \\[10pt]
	{\qquad\quad}\times {_4F_3}\bigg(\begin{array}{c} -p_1, \quad p_1 + \frac12, \quad  q_1 + \frac12, \quad -  q_1 \\
	 N + 1,\quad  1 , \quad  -  N \end{array}\bigg|\,1\bigg). \end{array} \lab{3step}
\ee

As a result of these three successive transformations we finally
arrive at the cano\-nical form of the Racah polynomials
\rf{Racah} in terms of the hypergeometric series $_4F_3$
in \rf{INT-03},
\be
	\begin{array}{l}
	\displaystyle{_4F_3}\bigg(\begin{array}{c} -q_2, \quad q_1 + N + 1, \quad q_1 + 1, \quad q_1 + 1\\
		 2 q_1 + \tsty32,\quad q_1 + p_1+ \tsty32, \quad q_1 - p_1 + 1 \end{array}\bigg|\,1\bigg)\\[10pt]
	\displaystyle{\qquad}= (-1)^{p_1}2^{2q_1+4p_1+1}
		\frac{\Gamma(q_1+ p_1  + \frac32)\,\Gamma(q_1 - p_1+1)}{\Gamma(p_1 + \frac12)}\\[10pt]
	\displaystyle{\quad\qquad}\times\frac {p_1!\,(2q_2)!\,(2p_2)!\,(q_1+N)!\,[(p_1+ N)!]^2 (4q_1+1)!
		}{q_2!\,[(p_2)!]^2\,[(2q_1)!]^2\, (2q_1 +2 N + 1)! \,(2p_1 +2 N + 1)!}\\[10pt]
	{\quad\qquad}\times
	\,{_4F_3}\bigg(\begin{array}{c} -p_1, \quad p_1 + \frac12, \quad  q_1 + \frac12, \quad -  q_1 \\
	 N + 1,\quad  1 , \quad  -  N \end{array}\bigg|\,1\bigg). \end{array} \lab{Finstep}
\ee

The parameters of $_4F_3$ hypergeometric functions 
for the interbasis coefficients 
$U_{2q_1+1, 2q_2}^{2p_1, 2p_2+1}$ in \rf{2-INT-04} 
and $U_{2q_1+1, 2q_2+1}^{2p_1+1, 2p_2+1}$ in \rf{2-INT-05} 
also enjoy the property \rf{Saal} and can 
be transformed into the canonical form for Racah polynomials
\rf{Racah} by using the three steps \rf{1step}-\rf{3step}
{\it mutatis mutandis}. This results in
\be
	\begin{array}{l}
	{\hskip-15pt}{_4F_3}\bigg(\begin{array}{c} -q_2, \quad  q_1 + N + 2, \quad  q_1 + 1, \quad  q_1 + 1\\
	2 q_1 + \tsty52,\quad  q_1 + p_1+ \tsty32, \quad  q_1 - p_1 + 1 \end{array}\bigg|\,1\bigg)\\[10pt]
	\displaystyle{}= (-1)^{p_1}\frac{}{}
		\frac{4^{q_1+2p_1+1}\,p_1!\,(2q_2)!\,(2p_2+1)!\,(q_1+N+1)!
		}{(N+1)\,q_2!\,[(p_2)!]^2\,[(2q_1+1)!]^2}\\[10pt]
	\displaystyle{}\times  \frac{\,[(p_1+ N+1)!]^2 (4q_1+3)!\,\Gamma(q_1+ p_1  + \tsty32)\,\Gamma(q_1 - p_1+1)
		}{(2q_1 +2 N + 3)! \,(2p_1 +2 N + 2)!\,\Gamma(p_1 + \tsty12)}\\[10pt]
	{\qquad}\times{_4F_3}\bigg(\begin{array}{c} -p_1, \quad  p_1 + \frac12, \quad   q_1 + \frac32, \quad  -  q_1 \\
		  N + 2,\quad   1 , \quad   -N \end{array}\bigg|\,1\bigg),
	\end{array} \lab{49'}
\ee
where $N= q_1 + q_2= p_1 + p_2=\onehalf(n-1)$, and
\be
	\begin{array}{l}  \displaystyle
	{\hskip-15pt}{_4F_3}\bigg(\begin{array}{c} -q_2, \quad  q_1 + N + 3, \quad  q_1 + 1, \quad  q_1 + 1\\
	 2 q_1 + \tsty52,\quad  q_1 + p_1+ \tsty52, \quad  q_1 - p_1 + 1 \end{array}\bigg|\,1\bigg) \\[10pt]
	\displaystyle{}= (-1)^{p_1} \frac{p_1!\,(2q_2+ 1)!\,(2p_2+1)!\,(q_1+N+1)!\,(p_1+ N+1)!
	}{(N+1)(N+2)\,q_2!\,[(p_2)!]^2\,[(2q_1+1)!]^2} \\[10pt]
	\displaystyle{}\times  \frac{4^{q_1+2p_1+1}\,(p_1 + N +2)!\,(4q_1+3)!\Gamma(q_1+ p_1  + \tsty52)\,\Gamma(q_1 - p_1+1)
			}{(2q_1 +2 N + 3)! \,(2p_1 +2 N + 3)!\,\Gamma(p_1 + \tsty32)}\\[10pt]
	{\qquad}\times{_4F_3}\bigg(\begin{array}{c} -p_1, \quad  p_1 + \frac32,\quad q_1 +\frac32,\quad -q_1 \\
		 N + 3,\quad   1 , \quad   -  N \end{array}\bigg|\,1\bigg),  \end{array}\lab{50'}
\ee
where $N= q_1 + q_2= p_1 + p_2=\onehalf n-1$.

\section*{Acknowledgements}

We thank Guillermo Kr\"otzsch ({\sc icf-unam}) for indispensable help
with the figures, and acknowledge the interest of Cristina
Salto-Alegre (PCF--UNAM). G.S.P.\ and A.Y.\ thank the support of project
{\sc pro-sni-2017} (Universidad de Guadalajara).
N.M.A.\ and K.B.W.\ acknowledge the support of {\sc unam-dgapa}
Project {\it \'Optica Matem\'atica\/} {\sc papiit}-IN101115.


\end{document}